\documentclass[%
reprint,
superscriptaddress,
amssymb,amsmath,
]
{revtex4-2}

\usepackage[english]{babel} 
\usepackage [table]{xcolor}

\usepackage{graphicx, upgreek} 

\usepackage{natbib}

\usepackage{float}
\usepackage{textcomp}
\usepackage{graphicx}
\usepackage{dcolumn}
\usepackage{bm}
\usepackage[mathlines]{lineno}

\usepackage{physics}
\usepackage{array}

\newcommand{\sitwoeight}{$^{28}$Si}
\newcommand{\ttwo}{$T_2$}

\begin{document}

\begin{abstract}
Ensembles of bismuth donor spins in silicon are promising storage elements for microwave quantum memories due to their long coherence times which exceed seconds. Operating an efficient quantum memory requires achieving critical coupling between the spin ensemble and a suitable high-quality factor resonator --- this in turn requires a thorough understanding of the lineshapes for the relevant spin resonance transitions, particularly considering the influence of the resonator itself on line broadening. 
Here, we present pulsed electron spin resonance measurements of ensembles of bismuth donors in natural silicon, above which niobium superconducting resonators have been patterned.
By studying spin transitions across a range of frequencies and fields we identify distinct line broadening mechanisms, and in particular those which can be suppressed by operating at magnetic-field-insensitive `clock transitions'. Given the donor concentrations and resonator used here, we measure a cooperativity $C\sim 0.2$ and based on our findings we discuss a route to achieve unit cooperativity, as required for a quantum memory. 
\end{abstract}

\title{Spin resonance linewidths of bismuth donors in silicon coupled to planar microresonators}

\author{James~O'Sullivan}
\altaffiliation{These authors have contributed equally to this work} 
\affiliation{London Centre for Nanotechnology, UCL, 17-19 Gordon Street, London, WC1H 0AH, UK}
\author{Oscar~W.~Kennedy}
\altaffiliation{These authors have contributed equally to this work} 
\affiliation{London Centre for Nanotechnology, UCL, 17-19 Gordon Street, London, WC1H 0AH, UK}
\author{Christoph~W.~Zollitsch}
\affiliation{London Centre for Nanotechnology, UCL, 17-19 Gordon Street, London, WC1H 0AH, UK}
\author{Mantas~\v{S}im\.{e}nas}
\affiliation{London Centre for Nanotechnology, UCL, 17-19 Gordon Street, London, WC1H 0AH, UK}
\author{Christopher~N.~Thomas}
\affiliation{Cavendish Laboratory, University of Cambridge, JJ Thomson Ave,  Cambridge CB3 0HE, UK}
\author{Leonid V. Abdurakhimov}
\altaffiliation[Present address: ]{NTT Basic Research Laboratories, NTT Corporation, 3-1 Morinosato-Wakamiya, Atsugi, Kanagawa 243-0198, Japan} 
\affiliation{London Centre for Nanotechnology, UCL, 17-19 Gordon Street, London, WC1H 0AH, UK} 
\altaffiliation{Present address: NTT Basic Research Laboratories, NTT Corporation, 3-1 Morinosato-Wakamiya, Atsugi, Kanagawa 243-0198, Japan}
\author{Stafford Withington}
\affiliation{Cavendish Laboratory, University of Cambridge, JJ Thomson Ave,  Cambridge CB3 0HE, UK}
\author{John~J.~L.~Morton}
\altaffiliation{jjl.morton@ucl.ac.uk} 
\affiliation{London Centre for Nanotechnology, UCL, 17-19 Gordon Street, London, WC1H 0AH, UK}
\affiliation{Department of Electrical and Electronic Engineering, UCL, Malet Place, London, WC1E 7JE, UK}

\maketitle

Despite their impressive role in the demonstration of `quantum supremacy'~\cite{arute2019quantum}, the coherence time of superconducting qubits remains shorter than many other solid state systems.  
Complementing such superconducting quantum processors with long lived microwave quantum memories could enhance their functionality, allowing quantum states to be stored and retrieved for further processing on much longer timescales.  
Ideal candidates for the storage element of quantum memories are electron spin systems which exhibit long coherence times and possess suitable transitions even in relatively low magnetic fields, to maximise compatibility with superconducting circuits and qubits. 
Bismuth donors in silicon have been identified as attractive candidates against such criteria~\cite{george2010electron}, possessing a number of `clock transitions' at microwave frequencies where the coupling of donor spins to each other, and to other sources of magnetic field noise, becomes heavily suppressed. At such points, coherence times of 0.1~s have been measured in natural silicon, rising to about 3~s in isotopically enriched \sitwoeight~\cite{, wolfowicz2013atomic}, over four orders of magnitude longer than that in state-of-the-art superconducting qubits.

Microwave quantum memories require a cavity (typically a planar superconducting resonator) either critically coupled \cite{afzelius2013proposal} or strongly coupled \cite{julsgaard2013quantum, grezes2016towards} to an ensemble of paramagnetic spins allowing for coherent exchange of quantum information. The coupling strength of the spin ensemble to the resonator, $g_{\rm ens}$, must be sufficient at least to achieve cooperativity $C = g_{\rm ens}^2/\kappa\gamma$ equal to 1, where $\kappa$ and $\gamma$ are the half width at half maximum (HWHM) of the cavity and spin lines, respectively.
In a cavity containing $N$ coherent spins with homogeneous single-spin coupling, $g_0$, the ensemble coupling strength, $g_{\rm ens}$ becomes enhanced through collective coupling: $g_{\rm ens} = g_0\sqrt{N}$, ~\cite{taviscummings}. 
A variety of paramagnetic spin ensembles with spin concentrations of 10$^{17-20}$cm$^{-3}$ have already been used to demonstrate high cooperativity coupling to planar superconducting resonators \cite{kubo2010strong, weichselbaumer2020echo, probst2013anisotropic,schuster2010high}. For example, using phosphorus donors in isotopically purified silicon $C\sim2$ was achieved, however, the coherence time, \ttwo, in that experiment was limited to only $\sim2$~ms, despite coherence times approaching seconds being measured in the same spin system in the dilute limit and in the bulk~\cite{ross2019electron}. The use of clock transitions in bismuth offers the prospect to achieve sufficient coupling while maintaining long coherence times --- for example a recent study of bismuth in \sitwoeight\ coupled to an aluminium resonator has shown \ttwo~$=0.3$~s combined with $C=3.5\times10^{-2}$~\cite{ranjan2020multimode}.

In addition to the ensemble coupling strength, which can be modified by the choice of resonator design and spin concentration, two further critical parameters are the resonator and electron spin resonance (ESR) linewidths. Superconducting resonators with quality factors above $10^6$ (corresponding to linewidths less than 10 kHz) have been demonstrated \cite{zollitsch2019tuning, megrant2012planar}. 
In contrast, ESR linewidths of bismuth donors in bulk-doped silicon have been measured at X-band ($\sim$9.7~GHz) to be 4.5~MHz in silicon with natural isotopic abundance ($^{\rm nat}$Si) and $\sim$ 0.1~MHz in ion implanted isotopically enriched \sitwoeight~ \cite{george2010electron,weis2012electrical}.  Thus, even for isotopically purified silicon it is the spin linewidth that determines the threshold of ensemble coupling strength required to enter the strong coupling regime and, if reduced, would increase cooperativity. Furthermore, strain-induced line broadening~\cite{mansir2018linear} has been observed for near-surface bismuth donors underneath aluminium resonators, showing ESR lineshapes dominated by inhomogeneous strain with linewidths $\sim7$~MHz even in \sitwoeight~\cite{pla2018strain}.

In this Article, we study ESR linewidths of bismuth donors in natural silicon, using niobium superconducting resonators, which are expected to induce less strain in the silicon due to a better matching of the thermal expansion coefficient. Furthermore, by examining a range of ESR transitions with different dependence of transition frequency (f) upon static external magnetic field strength ($B_0$), $\partial f/\partial B_0$, --- from $\sim28$~GHz/T, the gyromagnetic ratio of a free electron, down the limit where it tends to zero at a clock transition ---  we identify different line broadening mechanisms and expected limits of linewidths. Based on our results we establish how sufficiently strong coupling can be achieved between Bi donors and superconducting planar resonators to achieve a high-efficiency quantum memory.


\begin{figure}
    \centering
    \includegraphics[width=0.47\textwidth]{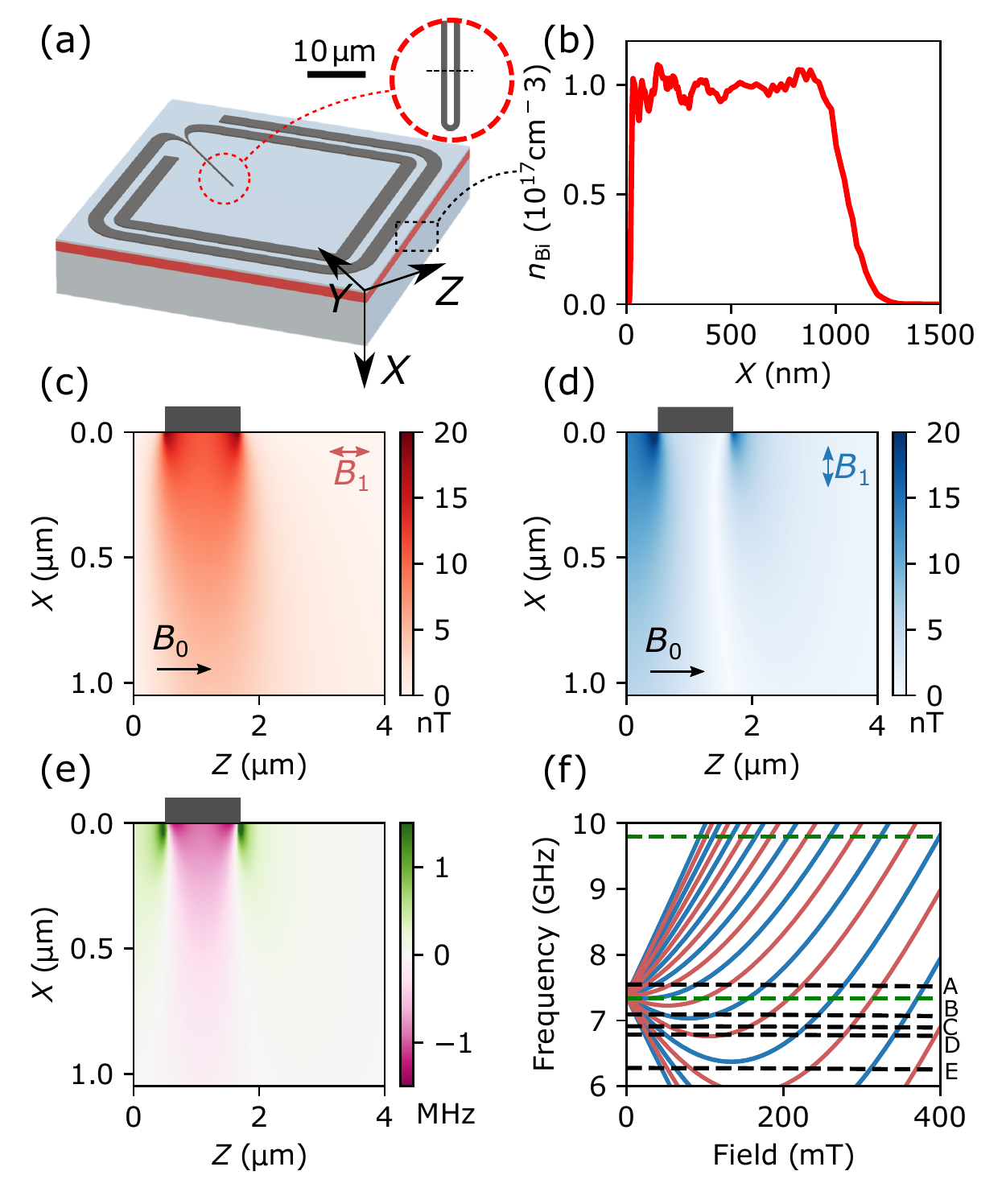}
    \caption{(a) Layout of the resonator with detail of inductor showing a dashed black line of the cross sections taken in (c,d). Axes are shown with X perpendicular to the substrate surface, Y along the hairpin inductor and Z perpendicular to the inductor. (b) Implantation profile of Bi calculated by Monte-Carlo simulations using the SRIM package \cite{ziegler2010srim}. (c, d) Two-dimensional finite element (FE) simulations of the microwave magnetic field $B_1$ due to zero-point-fluctuations of the resonator current assuming a 100~$\Omega$ characteristic impedance. (c) and (d) respectively show components which are parallel and perpendicular to the static magnetic field $B_0$ and correspondingly drive  $S_{\rm z}$ and $S_{\rm x}$ transitions. (e) FE simulations of shifts in Bi donor hyperfine constant due to strain due to the mismatch of thermal expansion coefficient between silicon and niobium. (f) Spectrum of $S_{\rm z}$ (red) and $S_{\rm x}$ (blue) spin transitions in Bi:Si as a function of magnetic field. Dashed lines indicate the frequencies of superconducting planar  (black) and non-superconducting 3D (green) resonators used in this work. Five superconducting resonators are labelled A-E with A (E) the highest (lowest) frequency resonator. There are several resonators at slightly higher frequencies than resonator C which are not shown here. }
    \label{fig:device}
\end{figure}

We fabricated resonators on a natural silicon substrate, using 100~nm thick sputtered niobium films, deposited using parameters that minimise film-induced strain in the substrate at room temperature and patterned using a standard lift-off process. The resonator design, shown in Fig.~\ref{fig:device}(a), contains a co-planar capacitor and an inductive hairpin where the conductor width is reduced and which is primarily responsible for coupling to spins. The hairpin geometry results in an anti-parallel microwave currents, confining magnetic fields close to the resonator and improving coupling efficiency to implanted layers of spins. Bismuth atoms were introduced into the silicon substrate, prior to resonator fabrication, by ion implantation into the top micron at a concentration of 10$^{17}$~cm$^{-3}$ (see Fig.~\ref{fig:device}(b)). Samples were annealed at 900$^\circ$C for 5~minutes to incorporate the bismuth atoms into the silicon lattice forming spin-active donors with an  efficiency of $\sim60\%$ \cite{peach2018effect}. We placed the resonators inside a copper 3D box in a setup similar to Ref.[~\citenum{bienfait2016reaching}] and measured them in transmission mode using two antennae. A full description of fabrication and the measurement setup is given in the Supplementary Information (SI)~\footnote{See Supplemental Material at [URL ] for additional experimental details.}.

The frequencies of the planar microresonators span a range from 6.28 to 7.55~GHz, intersecting with a number of bismuth donor ESR transitions, as shown in  Fig.~\ref{fig:device}(f).
The bismuth donor spin system comprises an electron spin $\mathbf{S}$ = 1/2 bound to the nuclear spin of $^{209}$Bi $\mathbf{I}$ = 9/2 and is described by the spin Hamiltonian in  frequency units:
\begin{equation}
\label{eq:ESR}
H/h= A\mathbf{I\cdot S}+\gamma_{\rm e}\mathbf{B_0\cdot S}+\gamma_{\rm Bi}\mathbf{B_0\cdot I} 
\end{equation}
where $A$ is the (isotropic) hyperfine coupling constant (1.475~GHz~\cite{feher1959electron,morley2010initialization}), 
and
$\mathbf{B_0}$ is the external magnetic field, while $\gamma_{\rm e} =  27.997(1)$~GHz/T and $\gamma_{\rm Bi} = 6.9(2)$~MHz/T  are respectively the gyromagnetic ratios of the bound electron and Bi nuclear spin~\cite{wolfowicz2013atomic}.
The spin eigenstates in the low field regime are best described by a total spin quantum number $\mathbf{F} = \mathbf{I} + \mathbf{S}$ and its projection onto $\mathbf{B_0}$, $m_F$. Two classes of ESR transition arise, 
referred to as $S_{\rm x}$ and $S_{\rm z}$, and are respectively driven by microwave fields $\mathbf{B_1}$ that are perpendicular and parallel to $\mathbf{B_0}$ (see Fig.~\ref{fig:device}(f)). Due to the different spatial distribution of such microwave fields (simulated given $\sim$50~nm penetration depth in Nb), as shown in Fig.~\ref{fig:device}(c, d), spins undergoing these different transitions are found in distinct physical locations around the inductor: $S_{\rm z}$ transitions couple most strongly for spins directly beneath the resonator inductor wires and $S_{\rm x}$ transitions for spins between the inductor wires.
The different transitions obey different selection rules where $\Delta F\Delta m_F=\pm1$ for $S_{\rm x}$ transitions and $\Delta m_F=0$ for $S_{\rm z}$. The $S_{\rm x}$ transitions come in pairs closely spaced in frequency with one transition corresponding to $\Delta F\Delta m_F=+1$ and another $\Delta F\Delta m_F=-1$. At most fields the separation of these transitions is less than the bismuth linewidth so they cannot be resolved. A more detailed analysis of the spin energy level diagram is given in the SI. 

\begin{figure}
    \centering
    \includegraphics[width=0.5\textwidth]{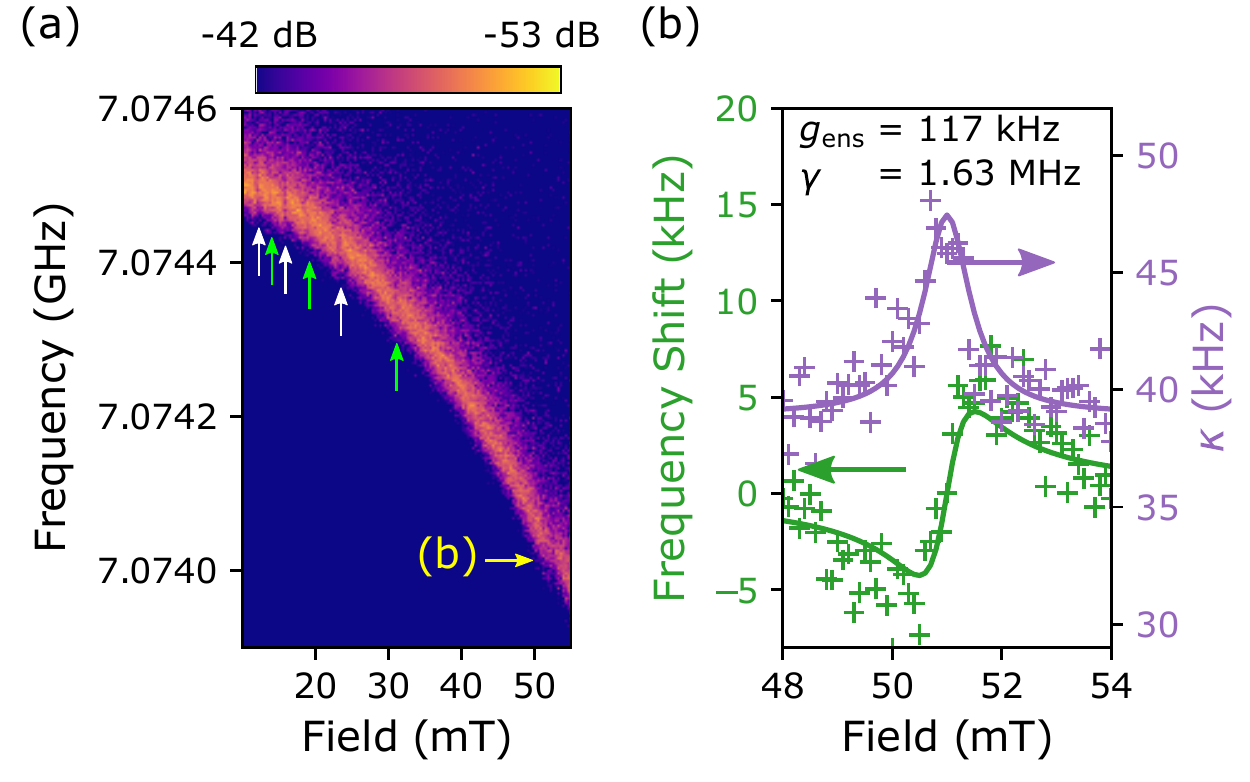}
    \caption{(a) Field dependent transmission $S_{21}$ through the cavity showing quadratic dependence of the frequency of resonator B on magnetic field. Interactions between spins and the resonator are highlighted with arrows where white (green) arrows identify $S_{\rm x}$ ($S_{\rm z}$) transitions. The $S_{\rm x}$ transition shown in (b) is labelled with a yellow arrow. (b) Fit to CW data of an $S_{\rm x}$ transition with $\partial f/\partial B_0~=~-0.11\gamma_e$ showing an ensemble coupling strength of 120~kHz and cooperativity of 0.2. The temperature is 140~mK.}
    \label{fig:CW}
\end{figure}

We first detected the ESR transitions in a continuous wave (CW) approach by measuring microwave transmission $S_{21}$ through the microresonator, using a vector network analyzer (VNA). Fig.~\ref{fig:CW}(a) shows both the quadratic decrease in resonator frequency with increasing magnetic field (due to the increase in kinetic inductance of the superconductor \cite{groll2010measurement, yip1992nonlinear}), as well as a series of avoided crossings where spin transitions couple to the resonator. 
By fitting the microwave transmission to a Fano resonance function~\cite{fano1961effects} at each magnetic field value, the shift in resonator frequency and reduction of quality factor caused by coupling to the spin transition can be observed (for example, see Fig.~\ref{fig:CW}(b)). 
When resonant with a spin transition $\kappa$, and the resonator frequency, $f$, are determined by the spin linewidth and ensemble coupling \cite{Eisuke2011electron}:
\begin{equation}
    \kappa = \kappa_0 + \frac{g_{\rm ens}^2\gamma}{\Delta^2 + \gamma^2}
    \label{eq:kappa}
\end{equation}
\begin{equation}
    f= f_0 - \frac{g_{\rm ens}^2\Delta}{\Delta^2 + \gamma^2}
    \label{eq:freq}
\end{equation}
where $\kappa_0$ and $f_0$ are respectively the half width and frequency of the resonator in the absence of spins, and the detuning $\Delta = (B_0 - B_R)\times \partial f/\partial B_0$, where $B_R$ is the magnetic field at which spins and resonator are resonant. The CW data therefore give a measure of the spin linewidth and also a measure of ensemble coupling strength --- for example, fits to the data shown in Fig.~\ref{fig:CW}(b) give $g_{\rm ens}=$120(10)~kHz and $\gamma=1.6(2)$~MHz. Together, these result in a cooperativity $C\sim0.2$ which, whilst the highest value reported thus far for bismuth donors, is a factor of five too small to achieve critical coupling. 

\begin{figure}
    \centering
    \includegraphics[width=0.5\textwidth]{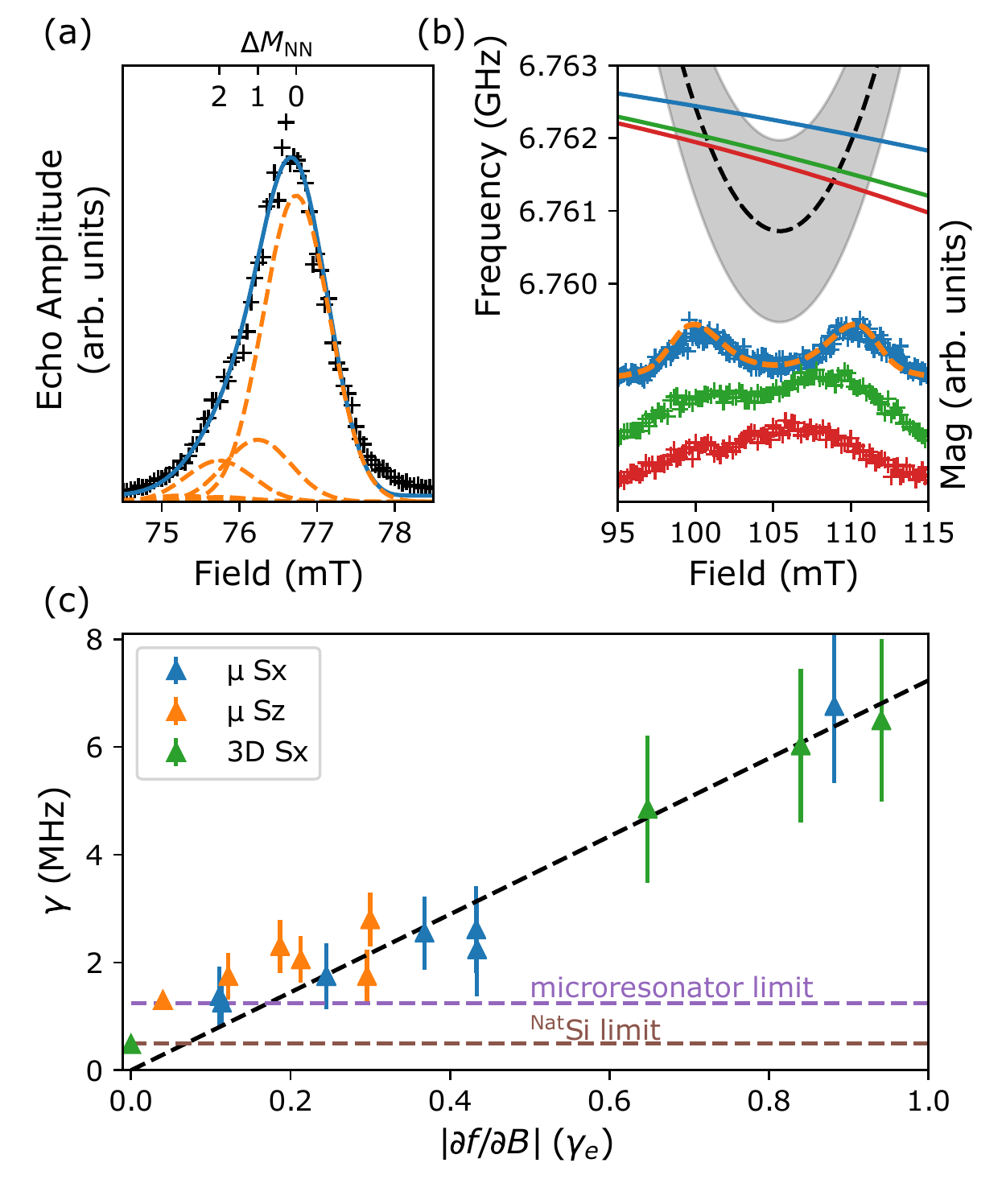}
    \caption{(a) EDFS of a resonator with slightly higher frequency than resonator C crossing an $S_{\rm z}$ transition with $\partial f/\partial B_0 = -0.12\gamma_e$  and a fit considering known effects from nearest-neighbour silicon isotopes. (b) EDFS at an $S_{\rm z}$ clock transition where resonator C has been tuned downwards showing well resolved split peaks which merge when the resonator is approaching clock transition. Solid lines show fits to the (weakly field-dependent) resonator frequency and dashed black line shows the spin transition with the 1.25~MHz full-width at half maximum of the inhomogeneously broadened line shaded in grey about the spin line. The dashed orange line overlaid on the data shows a fit to the EDFS considering nearest neighbour masses giving an intrinsic linewidth of 1.25~MHz. (c) Aggregate data (all resonators) showing width of individual Gaussians from fits as in (a, b) to transitions measured with different $\partial f/\partial B_0$ including $S_{\rm x}$ and $S_{\rm z}$ transitions measured using planar microresonators, as well as $S_{\rm x}$ transitions measured in a 3D cavity at X-band (9.66~GHz and 12~K) and a clock transition (7.3386~GHz and 10~K).
    CPMG-based averaging of 100 echoes was used to enhance the signal to noise ratio.}
    \label{fig:mass_fit}
\end{figure}

To measure the spin resonance lineshape in greater detail, the magnetic field was swept as pulsed ESR measurements were performed at the resonator frequency. A Carr-Purcell-Meiboom-Gill (CPMG) pulse sequence~\cite{Carr1954,Meiboom1958} was applied for the purposes of `echo-train averaging'~\cite{Mentink-Vigier2013}, and the emitted electron spin echo signal was amplified at multiple stages and captured on a digitizer (full details of the spectrometer and averaging scheme are shown in the SI).  The echo amplitude measured as a function of magnetic field in an echo-detected field sweep (EDFS) is shown for a representative $S_{\rm z}$ transition (6.8~GHz, $\partial f/\partial B_0 = -0.12\gamma_e$) in Fig.~\ref{fig:mass_fit}(a). These lineshapes are well described by a sum of Gaussians with centers determined by shifts to the hyperfine constant due to the average mass of nearest-neighbour silicon atoms weighted by the probability of each configuration of neighbours based on the natural isotopic abundance~\cite{sekiguchi2014host}. Given the relative amplitude and field shift for each of these additional peaks is fixed and known, we only need two free parameters for the fit: the central magnetic field and the common width of the constituent Gaussians (see SI for more details). The linewidths (half-width at half maximum) extracted in this way are shown in Fig.~\ref{fig:mass_fit} (c) for a number of $S_{\rm x}$ and $S_{\rm z}$ transitions, plotted as a function of the first-order magnetic field sensitivity, $\partial f/\partial B_0$ for each transition. The linear dependence of the linewidth upon $\partial f/\partial B_0$ implies $\sim$0.5~mT variations in effective magnetic field, arising from random distributions in $^{29}$Si nuclei.

Of particular interest is the limiting case where $\partial f/\partial B_0\sim0$, for example around the 6.76~GHz $S_{\rm z}$ `clock transition', where Fig.~\ref{fig:mass_fit}(b) shows the echo-detected magnetic field sweep
as resonator C is tuned downwards in frequency by tilting the applied field  with respect to the resonator plane~\cite{zollitsch2019tuning}. EDFS are collected at each $B_0$ as the resonator is tuned and are also shown in Fig.~\ref{fig:mass_fit}(b). When the resonator is at higher frequency (field perfectly aligned in the resonator plane), distinct peaks in echo amplitude on either side of the clock transition are clearly resolved. As the resonator frequency is tuned downwards, towards the clock transition, these peaks merge. As in Ref. [\citenum{wolfowicz2013atomic}], we can use the clock transition to precisely determine the value of hyperfine constant in this sample and find that A~=~1475.31~MHz, 140~kHz higher than measured in $^{28}$Si, attributed to the larger average mass of silicon atoms which is expected to increase the hyperfine constant by $\sim160$~kHz \cite{sekiguchi2014host}. In none of these spectra are individual nearest-neighbour mass-shifted peaks resolvable, implying that broadening of at least 0.85~MHz remains in these devices. A minimum linewidth of $\sim$1.25~MHz was inferred from the fits, which, based on the value of $\partial f/\partial A$ for this transition, equates to a HWHM in $A$ of 250~kHz, substantially greater than the value of 30~kHz measured in low-density bulk-doped Bi:$^{\rm nat}$Si at a clock transition~\cite{wolfowicz2013atomic}. 

In order to distinguish between potential contributions to the broadening seen here (such as ion-implantation damage, or strain from the microresonator), we measured linewidths in the ion-implanted sample using 3D microwave resonators (sapphire ring at 9.66~GHz and copper loop-gap resonator at 7.3386~GHz). Far from the clock transition the measurements from the 3D resonator match well the results from the microresonators, while at the 7.3~GHz clock transition a linewidth of $\sim$0.5~MHz was seen, equivalent to a HWHM in $A$ of 0.1~MHz, enabling Si isotope shifts to be resolved (see SI). The difference in the distribution of hyperfine couplings from bulk-doped Si and the ion-implanted samples studied here could be caused either by the 100$\times$ higher bismuth donor concentration and/or residual strain from ion implantation \cite{peach2018effect}. 
For spins in silicon beneath patterned microresonators, strain is known to arise from the different coefficients of thermal expansion (CTEs) of the various materials. While previous observations of microresonator-induced strain were based on aluminium~\cite{pla2018strain, bienfait2016reaching}, the niobium material used here has a CTE which is better matched to that of silicon. As a result, we observe shifts in hyperfine constant approximately half those seen for aluminium resonators~\cite{pla2018strain} despite these aluminium resonators having a minimum feature size four times larger than the resonators from this work. 

\begin{figure}
    \centering
    \includegraphics[width=0.5\textwidth]{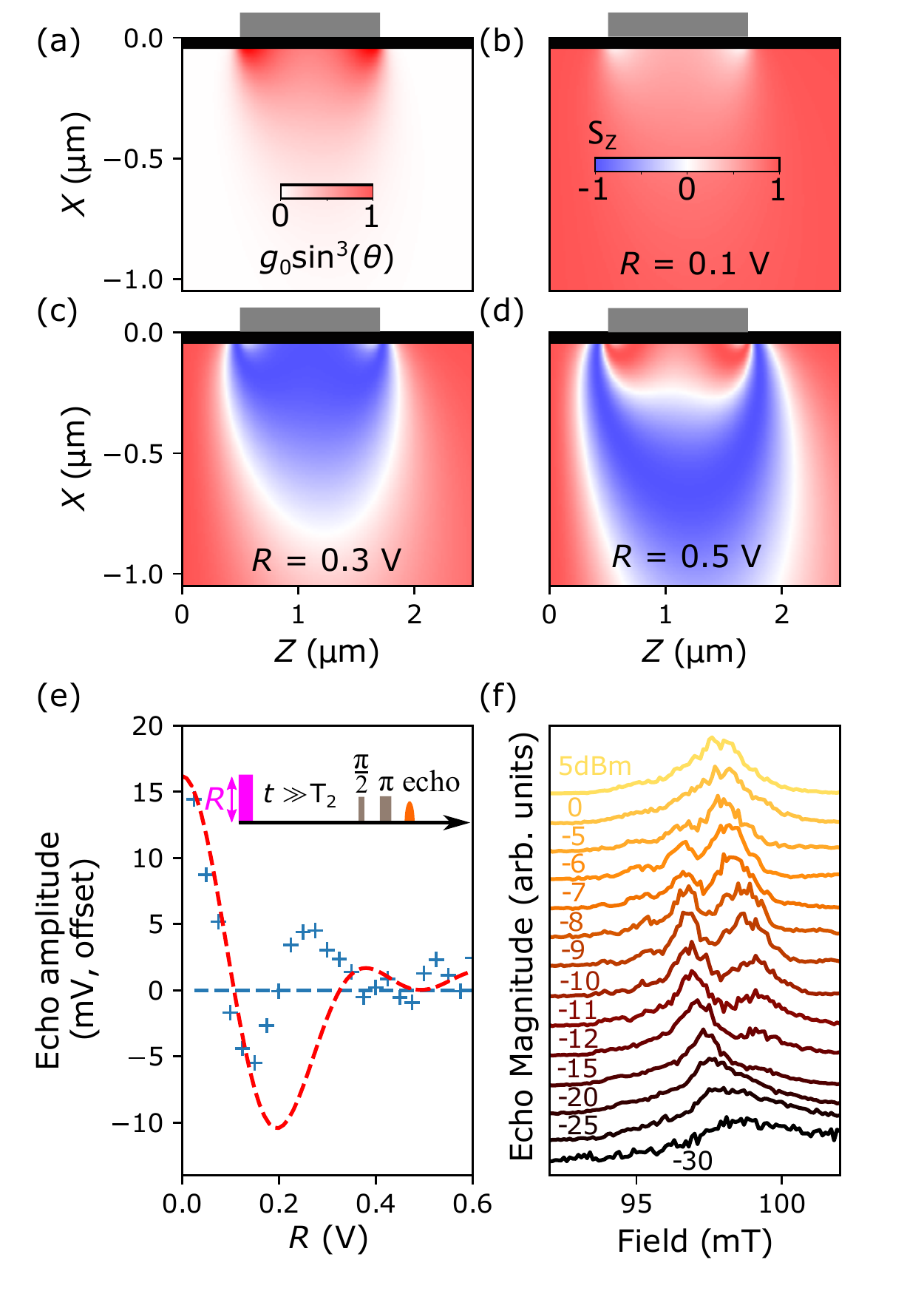}
\caption{Simulations of echo amplitude are performed by multiplying a contribution of each pixel (a) by the projection of the pixel along $z$ (b-d) and integrating over pixels. In (a-d) the resonator cross section is indicated in grey and donors assumed to be ionised due to band-bending are in black. 
(a) The calculated contribution of each pixel to a spin-echo measured using a two-pulse sequence ($\pi/2-\pi-$echo). The pulse amplitude was chosen to best match the data obtained using a power of about $-26$~dBm at the input antenna (estimated $\sim5\times10^5$ photons in the resonator---see SI \S G). 
(b–d) Simulated spatial variations of the $z$-projection of spins following a rectangular pulse of amplitude $R$ = 0.1, 0.3, 0.5~V. 
(e) Experimental (blue crosses) and simulated (red dashed line) Rabi oscillations collected on resonator C at the $S_{\rm z}$ clock transition shown in Fig.~3(b) (microwave power $-26$~dBm). The inset pulse sequence shows the first rotation pulse in magenta, the amplitude of which is swept in the Rabi experiment. 
(f) Power-dependent echo-detected field sweeps collected with a CPMG-1000 sequence on resonator C. 
}
    \label{fig:rabi_power}
\end{figure}

Due to the inhomogeneous coupling between spins and the microresonator, different spatial distributions of Bi spins can be addressed by varying the power used in pulsed ESR measurements~\cite{ranjan2020electron}. For example, measurements taken with lower microwave powers should be more sensitive to spins that are closer (and thus, more strongly coupled) to the resonator. In order to  investigate the spatial distribution of the ESR line broadening, we performed power-dependent pulsed ESR measurements and compared the results with 2D finite-element simulations of the strain distribution and expected ESR signals (see Fig.~\ref{fig:rabi_power}). Despite this $B_1$-inhomogeneity, uniform global manipulation of the spin ensemble remains possible using techniques such as  adiabatic fast-passage pulses could be employed to overcome this~\cite{kupce1995adiabatic, Sigillito2014}. 

We applied a 2~$\upmu$s pulse of variable amplitude followed some time $t = 10$~ms later (where $T_2 \ll t \ll T_1$) by a two-pulse Hahn echo detection sequence with pulses 2, 4~$\upmu$s respectively and $\tau=60~\upmu$s as shown in Fig.~\ref{fig:rabi_power}(e). In these samples $T_2$ varies depending on the transition and shot repetition time with all coherence lost after waits longer than 1~ms a detailed study of which is the subject of future work. $T_1$ varies between donors and is determined by the Purcell effect \cite{purcell1995spontaneous, bienfait2016controlling} with virtually all spins having $T_1>100$~ms. Using a low `detection' power for the Hahn echo sequence ($-26$~dBm at the source which is amplified and attenuated giving and $\sim-26$~dBm at the input antenna and further attenuated by insertion loss see SI for details) we observe damped Rabi oscillations as expected given the large range of microwave magnetic field strengths ($B_1$) experienced by different spins. 
The detection power determines the distribution of spins contributing to the observed signal, as illustrated in the 2D simulations shown in (Fig.~\ref{fig:rabi_power}(a)). The contribution of each pixel goes as $g_0 \sin^3(\theta(\mathbf{r}))$ \cite{rinard1999absolute} where $\theta(\mathbf{r}) = \gamma_eB_1(\mathbf{r})A_{\rm pulse}t_{\rm pulse}$ is the spatially varying tip angle induced by the first pulse in the Hahn echo sequence of amplitude $A_{\rm pulse}$ and duration $t_{\rm pulse}$.

Having selected a sub-ensemble of spins through the choice of `detection' power in the Hahn echo, we can then understand the effect of sweeping the amplitude of the initial pulse. The simulated projection $S_z$ of different spins following this initial pulse is shown in panels Fig.~\ref{fig:rabi_power}~(b-d) for different pulse amplitudes: at lower amplitudes only those spins closest to the resonator are flipped, while for larger amplitudes multiple bands of spins can be identified corresponding to spins having been rotated by even or odd multiples of $\pi$. Summing over all cells, weighted by their contribution to the detection Hahn echo signal (Fig.~\ref{fig:rabi_power}(a)) gives the simulated Rabi oscillations shown in Fig.~\ref{fig:rabi_power}(e). 
The qualitative agreement between the simulated and experimental pulsed ESR measurements at low detection powers suggests the simulations can be used to understand the microscopic distribution of spins contributing to the signal in a given ESR experiment, with spins closest to the resonator contributing most to experiments performed with lower power.

By varying the detection power we can obtain spin linewidth and lineshapes for different spatial distributions of spins \cite{ranjan2020pulsed}, as shown in Fig.~\ref{fig:rabi_power}(f). At the highest drive power of +5~dBm at the source giving $\sim5$~dBm at the input antenna (as used in Fig.~\ref{fig:mass_fit}) the echo contains contributions from many interfering bands of spins across a wide cross-sectional area of silicon beneath the resonator extending across the implanted spins (see SI, Fig.~S7, S8). Reducing the detection power reduces the distribution of spins and leads to narrower features in the echo-detected field-swept spectra. Oscillations in such spectra arise as different bands of spatially distributed spins, with distinct strain-shifted hyperfine constants, are tuned onto resonance at different magnetic fields. The half-width of these features is $\sim$0.5~MHz, similar to that seen in measurements using 3D cavities, confirming that narrower linewidths can be obtained by sampling a suitable sub-set of the spins to minimise the effect of strain-broadening and that magnetic field broadening is not increased in these devices relative to the 3D cavity measurement. 
At the lowest powers (probing spins closest to the resonator) the spectrum becomes a single broad peak --- although the spatial distribution is now very narrow, it corresponds to the region of highest strain, that immediately next to the resonator. Further details and simulations of these power-dependent magnetic field spectra are presented in the SI, along with arguments ruling out the Meissner effect as a contribution to line broadening for near-surface donors.    

A variety of approaches exist for increasing the cooperativity from the value of 0.2 measured here, to the levels required to achieve a bismuth donor spin microwave quantum memory. The first is simply to increase the density of implanted bismuth. While activation of the Bi donor spins can become challenging at donor densities above 3$\times$10$^{17}$~cm$^{-3}$, such a concentration would already provide sufficient increase in spin number to reach unit cooperativity that enables quantum memory through a  `critical coupling' protocol~\cite{afzelius2013proposal}. 
Additionally, the proportion of resonant bismuth donors can be increased by adjusting the nuclear spin populations from their thermal equilibrium values, for example, using resonant radiofrequency excitation~\cite{saeedi2013room} or optical pumping \cite{saeedi2013room, dreher2012nuclear}. Such an approach could increase the number of resonant spins by over twelve times the equilibrium value.

As well as increasing the ensemble coupling strength through the number of resonant spins, we consider techniques to reduce the linewidth of the bismuth donor spins. Using $^{ \rm Nat}$Si means that there are multiple
nearest-neighbor mass shifts to Bi hyperfine constant and only 72\% of
Bi have all four neighbors as 28Si. Using isotopically purified Si would concentrate all Bi donors into a single nearest-neighbour configuration and increase the resonant spin density by $\sim$40\%. Operating at a clock transition enables a narrowing of the linewidth of each nearest-neighbour peak down to a value of 1.25~MHz limited by strain, independent of isotopic purification. A reduction of the strain in the silicon directly beneath the patterned microresonators could be achieved either by control of sputtering parameters \cite{iosad1999optimization, glowacka2014development} or cryogenically cooling substrates during deposition. Eliminating strain in these devices could reduce the minimum spin linewidth by a factor of 2.5 (the difference in linewidth between patterned and un-patterned samples), and understanding the broadening mechanisms behind the linewidth of $\sim$0.5~MHz measured in a 3D cavity could lead to further improvements. Using isotopically purified silicon would remove nearest-neighbour mass shifts which also broaden the line. Each of the strategies above are likely to have an impact in the coherence time of the memory due to the impact in concentration of resonator donor spins.

Cooperativity is also determined by the loss rate from the resonator, given by the sum of $\kappa_i$, the internal loss rate of the resonator, and $\kappa_c$, the coupling loss rate of the resonator.
For spin echo experiments which form the basis of quantum memory protocols, we require $\kappa_c \gg \kappa_i$ such that when a spin echo forms, it can be coupled to a microwave bus rather than lost to the resonator environment. 
State-of-the-art planar microwave resonators have single photon quality factors in excess of 1 million \cite{megrant2012planar}, suggesting a target coupled resonator quality factor of 10k in order to achieve a memory efficiency of 0.99. In this work, the internal Q-factor was $\sim 100$k, with a coupled Q-factor of $\sim30$k, meaning that this device is sub-optimally coupled for a high efficiency quantum memory. In future devices higher internal quality factor resonators --- such as those already demonstrated at fields compatible with bismuth clock transitions \cite{zollitsch2019tuning, kroll2019magnetic} --- could give high efficiency (0.97) memories without impacting the cooperativity measured here. Cooperativity in the narrow $\kappa$ limit is discussed in the supplementary materials and Refs.[\citenum{grezes2015towards, julsgaard2012reflectivity}].

In conclusion, we measure the linewidth of spin transitions in Bi:Si in devices compatible with quantum memory protocols. We find that upon approaching clock transitions the linewidth of bismuth donor spin transitions narrows to 1.25~MHz for superconducting planar microresonators patterned upon the doped silicon and 0.5~MHz in unpatterned devices. We show that the linewidth limit in microresonators is caused by strain at the metal/silicon interface and, based on our results, map out a route to achieving sufficiently strong coupling to develop microwave quantum memories based on bismuth donor spins in silicon. 

\section{Acknowledgements}
The authors acknowledge the UK National Ion Beam Centre (UKNIBC) where the silicon samples were ion implanted and Nianhua Peng who performed the ion implantation. This project has received funding from the U.K. Engineering and Physical Sciences Research Council, through UCLQ postdoctoral fellowships (O.~W.~K, M.~S.) Grant Number EP/P510270/1 and a Doctoral Training Grant (J.~O'S). JJLM acknowledges funding from the European Research Council under the European Union's Horizon 2020 research and innovation programme (Grant agreement No. 771493 (LOQO-MOTIONS).
\bibliography{bibliography}

\end{document}


\beginsupplement

\title{Supplementary Information: Spin resonance linewidths of bismuth donors in silicon coupled to planar microresonators}

\author{James~O'Sullivan}
\altaffiliation{These authors have contributed equally to this work} 
\affiliation{London Centre for Nanotechnology, UCL, 17-19 Gordon Street, London, WC1H 0AH, UK}
\author{Oscar~W.~Kennedy}
\altaffiliation{These authors have contributed equally to this work} 
\affiliation{London Centre for Nanotechnology, UCL, 17-19 Gordon Street, London, WC1H 0AH, UK}
\author{Christoph~W.~Zollitsch}
\affiliation{London Centre for Nanotechnology, UCL, 17-19 Gordon Street, London, WC1H 0AH, UK}
\author{Mantas~\v{S}im\.{e}nas}
\affiliation{London Centre for Nanotechnology, UCL, 17-19 Gordon Street, London, WC1H 0AH, UK}
\author{Christopher~N.~Thomas}
\affiliation{Cavendish Laboratory, University of Cambridge, JJ Thomson Ave,  Cambridge CB3 0HE, UK}
\author{Leonid V. Abdurakhimov}
\altaffiliation[Present address: ]{NTT Basic Research Laboratories, NTT Corporation, 3-1 Morinosato-Wakamiya, Atsugi, Kanagawa 243-0198, Japan} 
\affiliation{London Centre for Nanotechnology, UCL, 17-19 Gordon Street, London, WC1H 0AH, UK} 
\altaffiliation{Present address: NTT Basic Research Laboratories, NTT Corporation, 3-1 Morinosato-Wakamiya, Atsugi, Kanagawa 243-0198, Japan}
\author{Stafford Withington}
\affiliation{Cavendish Laboratory, University of Cambridge, JJ Thomson Ave,  Cambridge CB3 0HE, UK}
\author{John~J.~L.~Morton}
\altaffiliation{jjl.morton@ucl.ac.uk} 
\affiliation{London Centre for Nanotechnology, UCL, 17-19 Gordon Street, London, WC1H 0AH, UK}
\affiliation{Department of Electrical and Electronic Engineering, UCL, Malet Place, London, WC1E 7JE, UK}

\maketitle

\subsection{The bismuth donor spin system} \label{subsec:bismuth_physics}
\begin{figure}[!htb]
	\includegraphics[width = \linewidth]{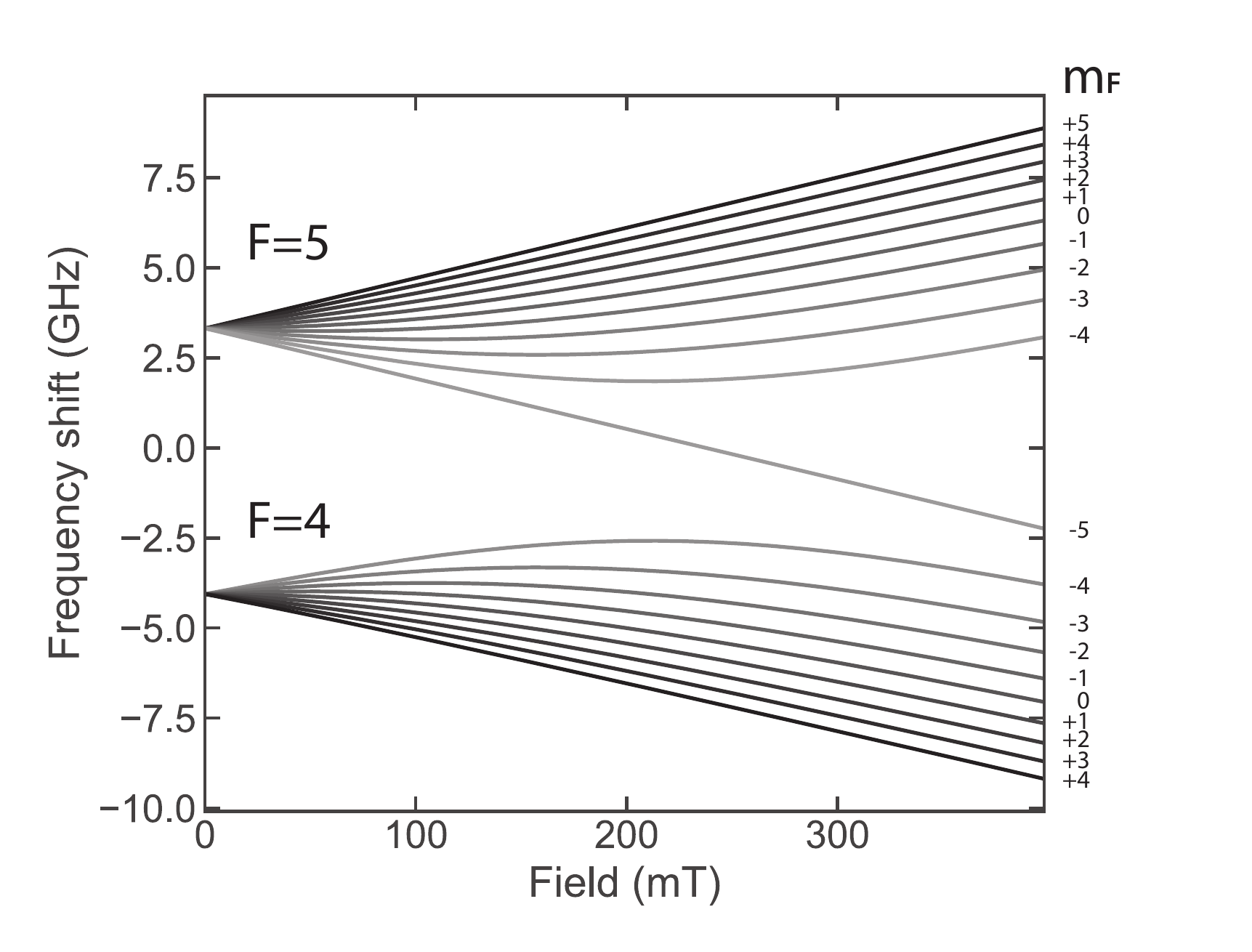}
	\caption[Breit-Rabi diagram of bismuth spin system]{Breit-Rabi diagram of Bi:Si spin system. At fields below $\sim$100~mT, the hyperfine interaction dominates, hence we label states by their total spin state $F$ and its projection on the $z$ axis $m_F$. Although from $100-400~$mT we are in a regime where the Zeeman and hyperfine interactions are similar in strength, we will continue to use the low field quantum numbers $\mathbf{F}$ and $m_F$ for simplicity.\label{fig:bismuth_breitrabi} }
\end{figure}

The Bi:Si spin Hamiltonian results, under the application of a magnetic field, in 20 non-degenerate energy levels, as shown in the Breit-Rabi diagram in Fig.~\ref{fig:bismuth_breitrabi}. At low fields, we label the eigenstates by the total spin $\mathbf{F}$ and its projection onto the $z$ axis $m_F$.

We now wish to find the relative transition strengths (given by the transition rates) between any two eigenstates under the influence of an arbitrary drive field $B_1$. To first order, the transition rate $\Gamma$ from one initial state $\left|\left.\ i\right\rangle\right.$ into a continuum of final states $\left|\left.\ f\right\rangle\right.$ with a density of final states $\rho$ of a system under the influence of a perturbing Hamiltonian $H$ is governed by Fermi's Golden Rule (FGR):
\begin{equation}
\Gamma=\frac{2\pi}{\hbar}\left|\left\langle\ f\middle|\ H\middle|\ i\right\rangle\right|^2\rho.
\end{equation}
In this case we have a single discrete final state, thus the density of states can be considered to be a delta function at the frequency of the transition. As we are interested only in the relative transition rates we can discard all constants and use a normalised field vector $\hat{\mathbf{B}}$ as the perturbing field:

\begin{equation}
\Gamma \propto \left|\left\langle\ f\middle|\ \hat{\mathbf{B}}\cdot\mathbf{S}\middle|\ i\right\rangle\right|^2.
\end{equation}

This allows us to determine the relative transition strengths and predicts two types of transition, $S_{\rm x}$ and $S_{\rm z}$, corresponding to $\hat{\mathbf{B}} = (1,0,0)$ and $\hat{\mathbf{B}} = (0,0,1)$, respectively.

\subsection{Measurement Setup and device fabrication and characterization}

\begin{figure}[!htb]
	\includegraphics[width = \linewidth]{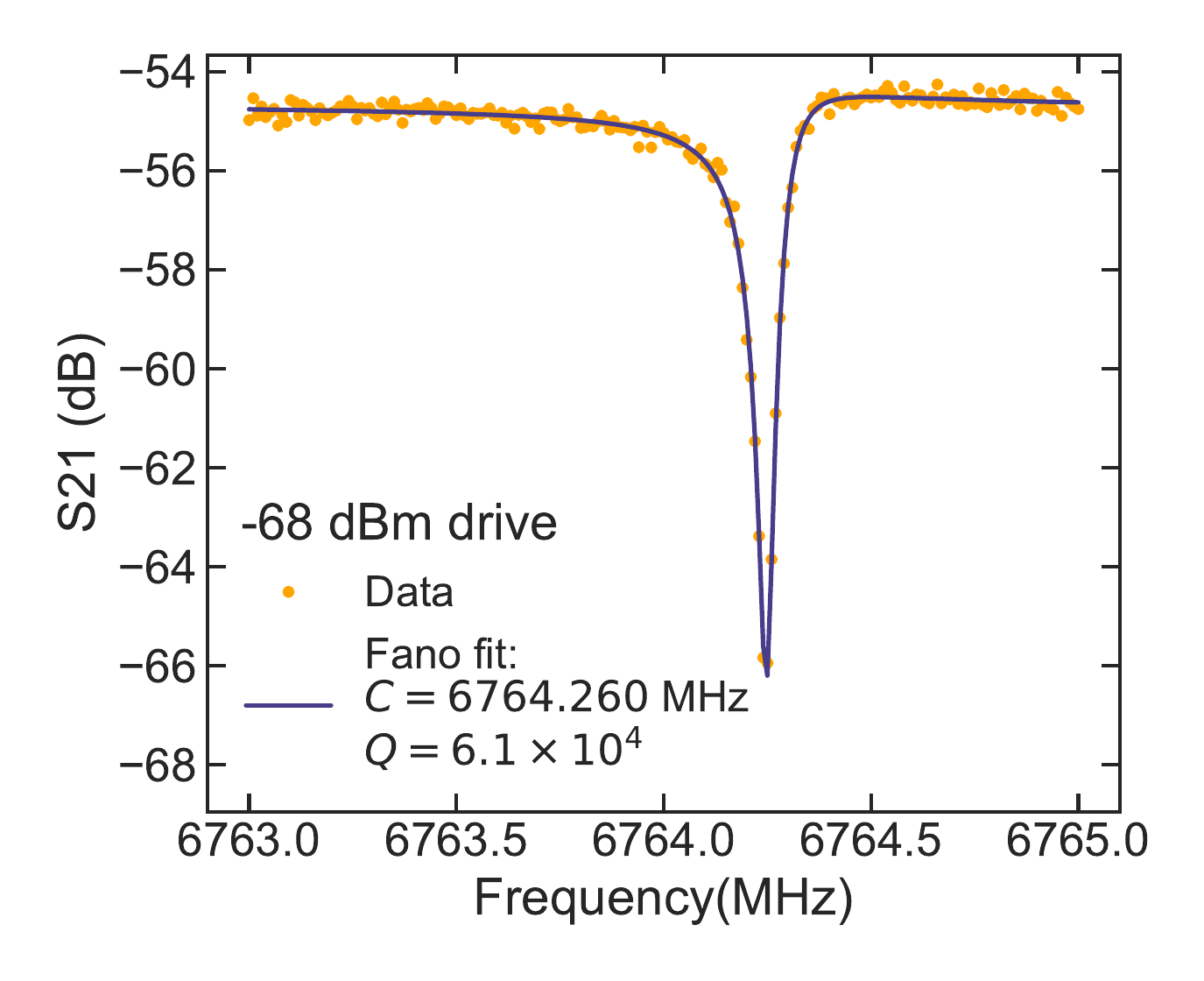}
	\caption{$S_{21}$ measurement of a resonator at -68 dBm at the antenna of the copper box. Data (yellow) is overlaid by a fit to Eqn.~\ref{eq:fano} of the SI (blue) giving a resonator frequency of 6764.260 MHz and Q-factor of 6.1$\times10^4$. The excellent fit to theory means that the fit precisely overlays the data in the centre of the resonance.\label{fig:fanofit }}
\end{figure}


The resonators were patterned onto the Bi implanted Si using a lift-off process.  For this process, the Si was diced into 13.5\,mm $\times$ 7.5\,mm chips for compatibility with tooling.  Several device chips were fabricated simultaneously on each of the larger host chips.  The photoresist stencil was formed by an image reversal process using optical contact lithography.  This process was optimized to give an undercut resist profile, which helps to eliminate fence formation at the film edges on lift-off.  The 100~nm of niobium was deposited by DC sputtering in a system with a base pressure of 5 $\times$ 10$^{-10}$\,mTorr. Sputtering power was 129\,W and Ar pressure 3.4\,mTorr.  Deposition time to give 100~nm was determined by calibration.  The deposition conditions had been optimized previously to give a film with low intrinsic stress on Si-like substrates.  For 100~nm niobium thick films deposited in this way, we typically measure a resistivity of 8.8\,$\upmu\Omega$cm at 20\,K and superconducting transition temperature in the range 8-9\,K. 

To characterize the microresonators we measure $S_{21}$ transmission through a copper box which is modulated by the microresonator. We fit the modulation (in linear magnitude) with a Breit-Wigner-Fano function:
\begin{equation}
    S_{\rm 21 Lin}(f) = K\frac{q\kappa + f - f_0}{\kappa^2 + (f - f_0)^2} + mx + c
    \label{eq:fano}
\end{equation}
where $K$ determines the size of the modulation, $q$ is an asymmetry parameter, $f_0$ is the central frequency of the resonator, $\kappa$ is the HWHM of the resonator and the $mx+c$ term is an approximation to the background transmission. The Fano resonance lineshape arises as a result of interference of microwave signals travelling directly from the input antenna to the recieve antenna with the signal absorbed and remitted by the resonator. In the limit of zero direct transmission between the antennae, this function reduces to a Lorentzian, which is the underlying true resonance lineshape. This function fits our resonance notch well and an example is shown in Fig.~\ref{fig:fanofit }.


\begin{figure}[!htb]
	\includegraphics[width = 0.8\linewidth]{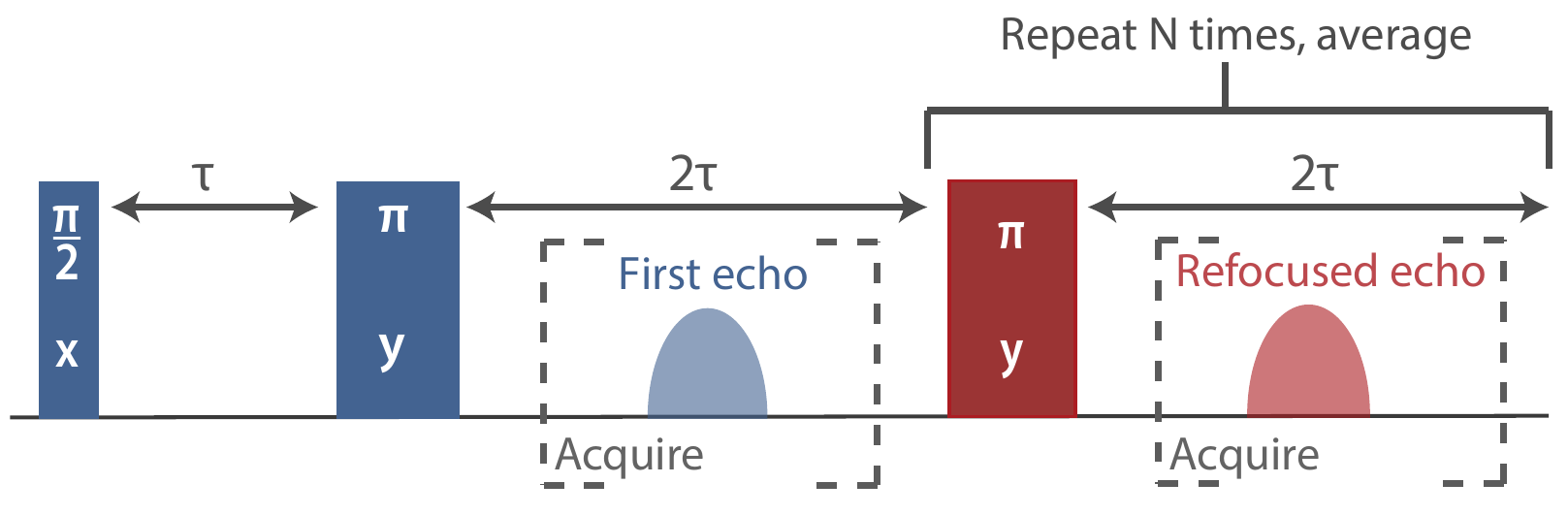}
	\caption{Schematic of the CPMG averaging scheme. After the initial excitation and refocusing pulses, spaced by time $\tau$, a train of $N$ refocusing pulses is applied after the first echo with spacing $2\tau$, generating a train of echoes. These are all acquired in a single shot and averaged. \label{fig:CPMG} }
\end{figure}

\begin{figure}
    \centering
    \includegraphics[width=0.5\textwidth]{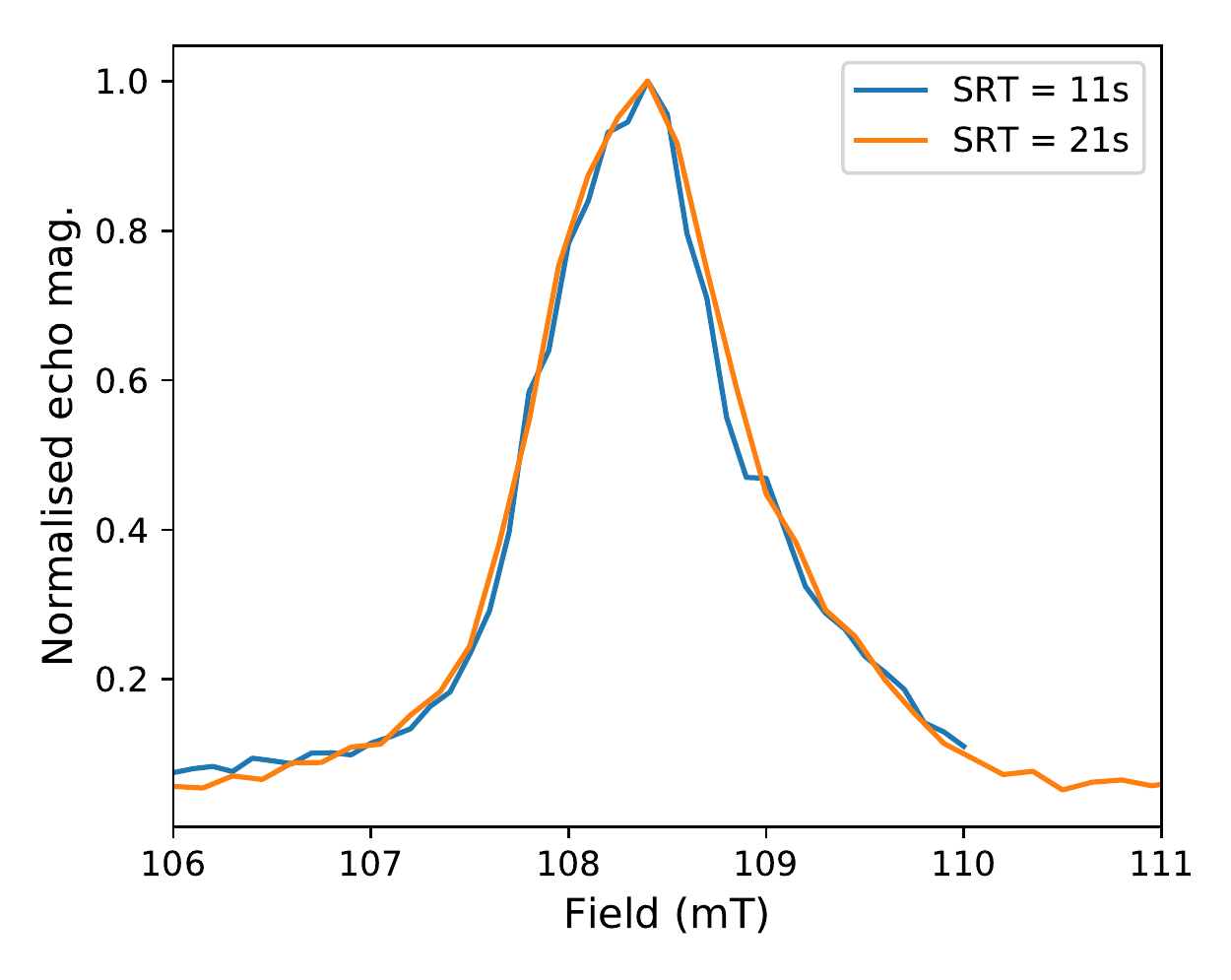}
    \caption{Normalised CPMG EDFS collected at 11s and 21s SRT showing no dependence of the linshape on SRT.}
    \label{fig:SRT}
\end{figure}

A technique for reducing measurement time, shown in Fig.~\ref{fig:CPMG}, was employed for Hahn echo detection for these measurements. After the Hahn echo has been measured, it is possible to refocus the spin system again to form another echo using a CPMG pulse sequence \cite{Carr1954,Meiboom1958}, provided the coherence time and $T_1$ relaxation time is sufficiently long. This is because only a small fraction of the energy stored in the spins is emitted during a Hahn echo. It has been shown by Mentink-Vigier et al. that such pulse sequences could be used for the purposes of averaging to improve ESR sensitivity~\cite{Mentink-Vigier2013}. In these measurements the $\pi/2$ ($\pi$) pulses were 2~$\upmu$s (4~$\upmu$s) and $\tau = 60~\upmu$s. In field sweeps, typically the field step results in a frequency shift greater than the resonator bandwidth and take $\sim$30~s for each field step which dominates the shot repetition time (SRT). Extensive tests show that, as expected, lineshapes taken with CPMG acquisitions are insensitive to SRT as shown, for example, in Fig.~\ref{fig:SRT}.

\begin{figure*}[!htb]
	\includegraphics[width = 0.7\textwidth]{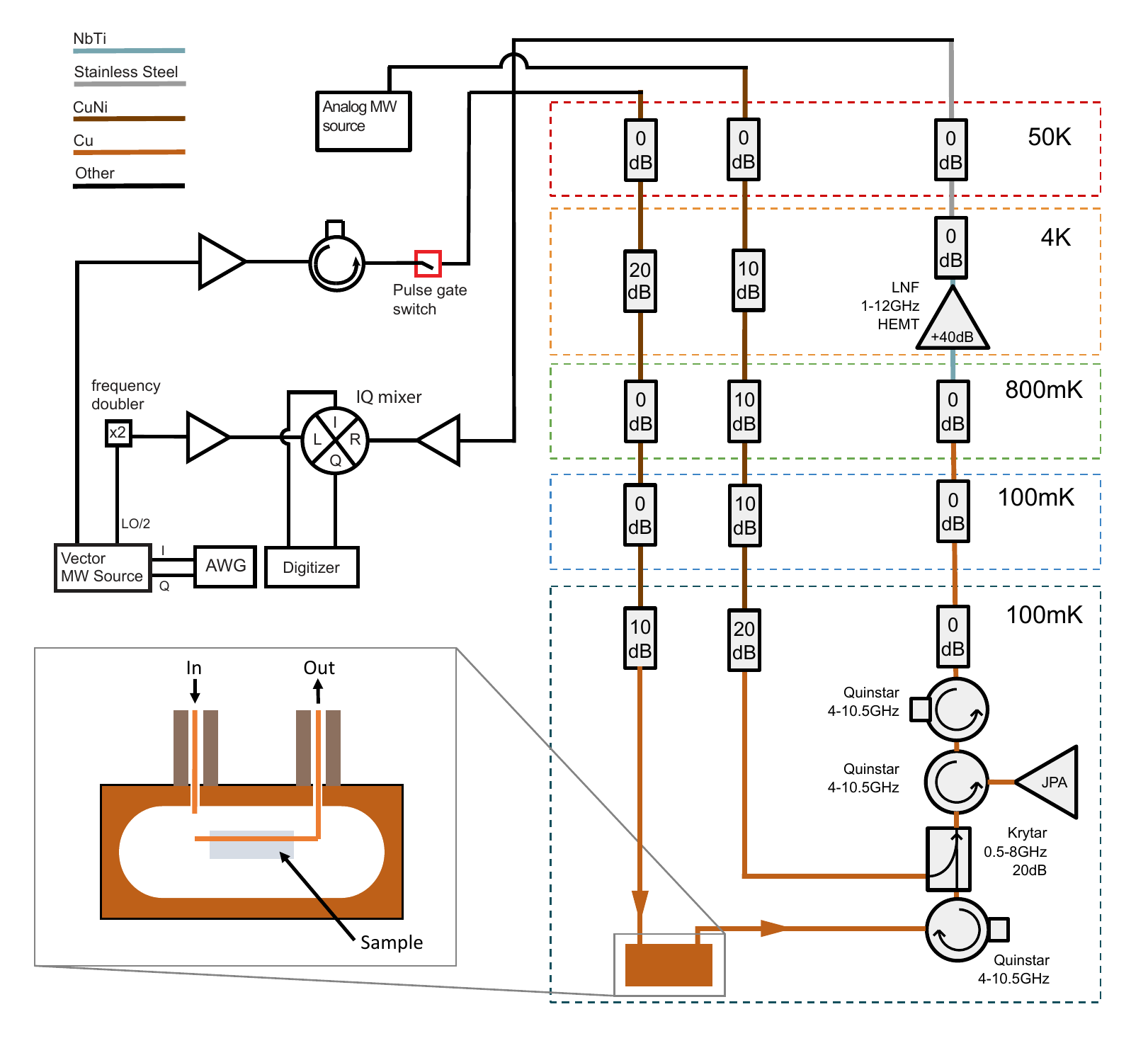}
	\caption{Schematic of the spectrometer and dilution fridge setup. The copper box housing the resonator is magnified to show the configuration of the sample and antennae inside. The receive antenna is considerably longer than the input antenna to increase coupling to the resonator and maximise signal. Coaxial cable types inside the fridge have been colour coded.\label{fig:Setup_full} }
\end{figure*}

The full measurement setup is shown in Fig.~\ref{fig:Setup_full}. An arbitrary wave-form generator (AWG) is used to send pulse sequences to a vector signal generator (VSG) which is set to the frequency of the superconducting resonator as determined by fitting to a VNA transmission measurement of the resonator. The pulse emitted from the VSG is amplified using a 30dB solid state amplifier with max output power +35~dBm before passing through cryogenic attenuators (30~dB in total) to thermalise the centre conductor of the input line and attenuate room temperature noise. Fast microwave switches gate the signal from the pulse amplifier when pulses are not being sent to ensure amplified room temperature noise is not sent into the fridge. The signal from the receive antenna inside the copper box is sent via two circulators to a Josephson Parametric Amplifier (JPA) which amplifies the signal in reflection. The amplifier is driven via a directional coupler by a dedicated microwave source. The JPA has typical gain of 10~dB although this varies depending on the settings of the JPA which is optimized for signal to noise ratio, rather than absolute gain. The signal is further amplifed by $\sim$40~dB at 4~K by a high electron mobility transistor (HEMT) and again by $\sim$40dB at room temperature before being mixed down in frequency by an IQ mixer. The IQ mixer uses a local oscillator (LO) produced by the VSG to mix down the signal. The VSG outputs a reference signal at half the VSG signal frequency and so a frequency doubler is used to produce a LO at the correct frequency which is amplified by 20dB to produce a signal with sufficient power for the IQ mixer. The mixed down signal is detected using a digitiser.

\subsection{Nearest neighbour mass fit}
\begin{figure}
    \centering
    \includegraphics[width=0.5\textwidth]{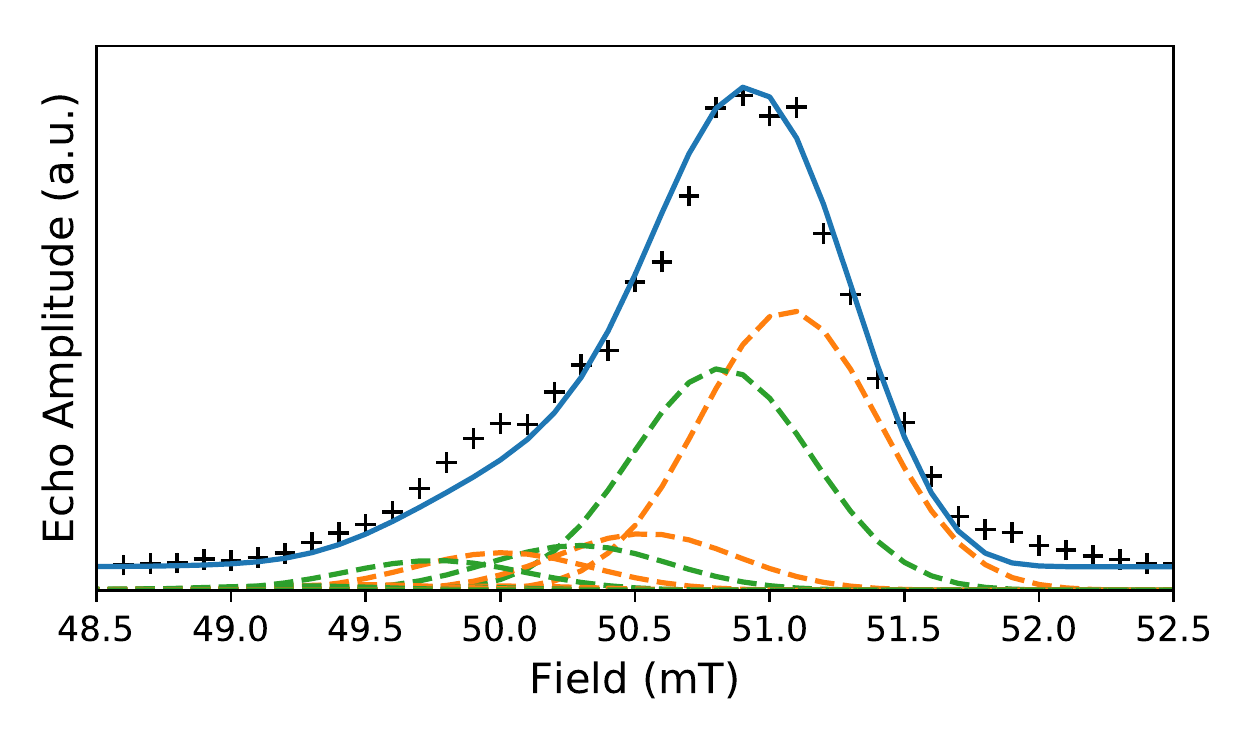}
    \caption{A fit to an EDFS of an $S_{\rm x}$ transition considering nearest neighbour mass shifts and quasi-degenerate transitions. The peaks from the two transitions are shown in green and orange, with relative amplitude determined by the FGR matrix element. The total fit is shown in blue and data as black crosses.}
    \label{fig:Sxfit}
\end{figure}

To fit echo detected field sweeps (EDFS) we consider the shifts to hyperfine constant due to the average mass of the bismuth donor's nearest neighbour silicon atoms. Using the isotopic abundance of natural silicon, we calculate a trinomial distribution to determine the probability of the different nearest neighbour configurations. At $\partial f/\partial A = 5$, as all of these transitions are up to a few $\%$, the shift is 1.7~MHz/$\Delta M_{\rm NN}$ where $\Delta M_{\rm NN}$ is the difference between the nuclear mass of all four nearest neighbour Si atoms and four $^{28}$Si. This means we know both the fraction of bismuth donors in each configuration and the relative frequency of donors in each configuration. We define a series of Gaussians with a common width and with relative amplitude determined by the fraction of bismuth donors in the configuration and relative frequency based on $\Delta M_{\rm NN}$ for the configuration. For $S_{\rm x}$ transitions where transitions are quasi-degenerate we consider the two transitions, calculating an offset frequency from the spin Hamiltonian and relative amplitudes from the matrix element in FGR as shown in Fig.~\ref{fig:Sxfit}. 

We then fit the EDFS to the series of Gaussians with two free parameters, the common width of the Gaussians and a correction to the field where the $\Delta M_{\rm NN} = 0$ line intersects with the resonator to allow for small field miscalibrations which are always $<1\%$. This results in fits which accurately capture the lineshape such as that shown in Fig.~3(a) in the main text. A tail at high frequency (which for the transition with negative $\partial f/\partial B$ shown in Fig.~~3(a) is at high field) which is badly captured by the fit routine is common to many transitions but more clearly resolved for transitions with low $\abs{\partial f/\partial B}$. This may indicate some regions of high strain.

\subsection{Clock transition measurements with 3D cavity}
\begin{figure*}
    \centering
    \includegraphics[width=\textwidth]{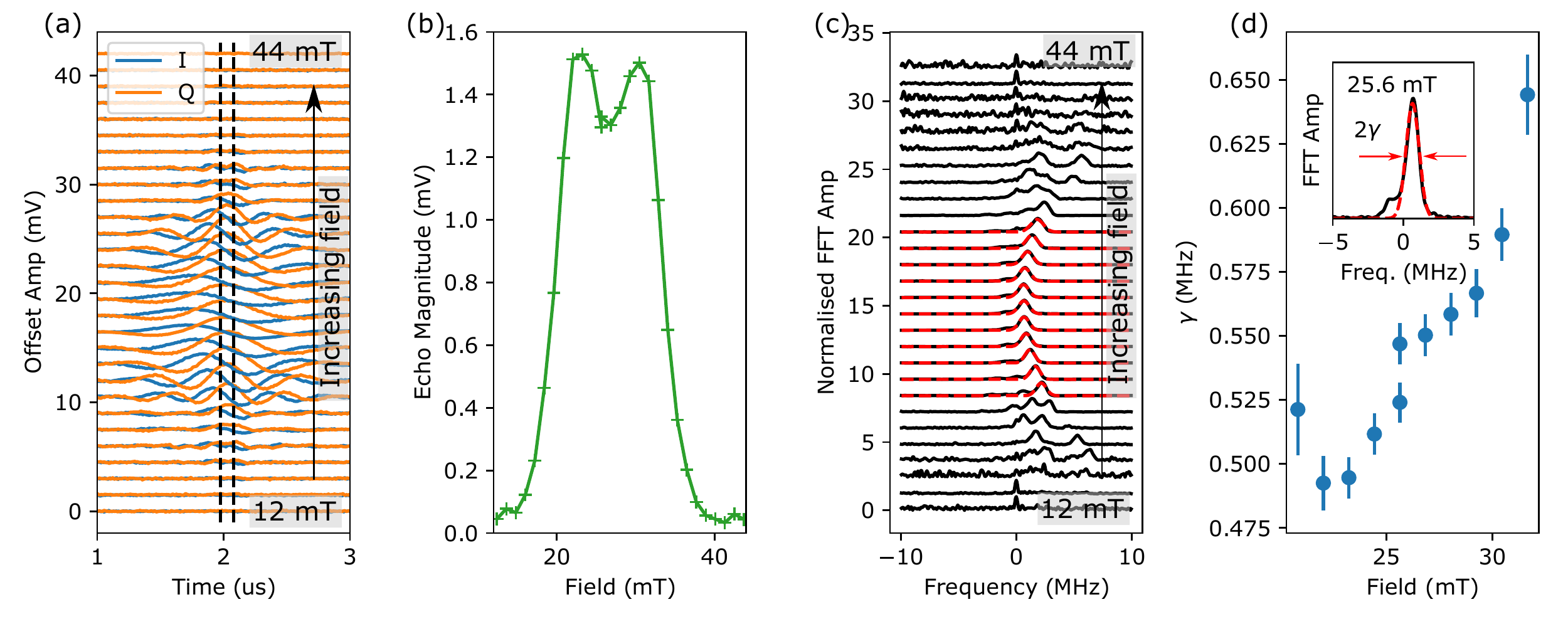}
    \caption{Results of EDFS measurements taken at an $S_{\rm x}$ transition using a 3D loop-gap copper resonator at 10~K. (a) Raw echo signal quadratures I (blue curves) and Q (orange curves) at different field values swept across the clock transition. Dashed vertical lines correspond to the integration time interval which was used to calculate data shown in Fig.\ref{fig:FFT_echo}(c). (c) Black curves correspond to normalised amplitudes of fast Fourier transform (FFT) of the complex data, $I + j\times Q$, where I and Q are signal components shown in Fig.\ref{fig:FFT_echo}(a).In addition to the dominant peaks, weaker peaks from nearest-neighbour mass shifts are resolved. The clock transition, where dominant peaks reach a minimum in frequency, is clear. Red curves correspond to results of a Gaussian fitting of the dominant peaks at the clock transition. (d) Spin linewidths $\gamma$ (half-width-at-half-maximum, HWHM) extracted from the data shown in Fig.\ref{fig:FFT_echo}(c) using the Gaussian fitting of the dominant peaks. The minimum linewidth value is 0.5~MHz. The inset shows an example of fitting results at the centre of the clock transition (also shown in Fig.\ref{fig:FFT_echo}(c))}
    \label{fig:FFT_echo}
\end{figure*}


Bismuth donors are measured at a clock transition using a 3D cavity. This means that no planar devices with mismatched CTE are present on top of the host Si and thus it remains unstrained. Using the same loop-gap copper resonator as that used in Ref.~\citenum{wolfowicz2013atomic} we measure an EDFS at the 7.38~GHz clock transition. Due to the small number of spins in a thin implanted layer we use a modified ESR probe with an in-built cryogenic HEMT amplifier described in a forthcoming publication. The results of this are shown in Fig.~\ref{fig:FFT_echo}. 

In Fig.~\ref{fig:FFT_echo}(a) we show the echoes in the time domain as the field is stepped across the clock transition in 1~mT steps. Integrating the centre of the echo magnitude as indicated by the dashed black lines in Fig.~\ref{fig:FFT_echo}(a) we obtain an EDFS shown in Fig.~\ref{fig:FFT_echo}(b). The dip at the centre of the clock transition is caused by the $\sim$10~\% of bismuth donors with nearest neighbour mass shifts of 3.4~MHz (i.e. $\Delta M_{\rm NN} = 2$) dropping out of the bandwidth of the cavity. Performing the fast fourier transform (FFT) of the echoes shown in Fig.~\ref{fig:FFT_echo}(a) allows us to resolve the spectrum of the bismuth. The parabolic dispersion of the most intense peak which is due to all nearest neighbour silicon atoms being $^{28}$Si is clearly seen. These peaks are fit by a Gaussian shown in red at the bottom of the clock transition and their width is shown in Fig.~\ref{fig:FFT_echo}(d). In addition to the peaks arising from nearest neighbours all being $^{28}$Si we can resolve peaks arising from $\Delta M_{\rm NN} = 1,2$. These peaks are weaker but follow the same quadratic dispersion at the bottom of the clock transition. We extract the minimum linewidth of the bismuth donors by taking the HWHM of the Gaussin which fits the FFT in Fig.~\ref{fig:FFT_echo}(c) and show linewidth is minimized at $\sim$0.5~MHz. 

\subsection{Resonator and strain simulations}
Resonators are simulated using COMSOL~\texttrademark~with simulations of the magnetic field distribution about the resonator and strain induced by differences in coefficients in thermal expansion solved together. 

The magnetic field about the resonator is determined following Refs. [\citenum{pla2018strain, bienfait2016reaching}]. The integrated current due to zero point fluctuations (ZPF), $\delta i = 2\pi f\sqrt{\hbar/2Z}$ where $Z$ is the impedance of the resonator. The impedance is determined assuming a lumped element geometry where $Z = \sqrt{L/C}$ and $f = 2\pi\sqrt{1/LC}$. The capacitance of the capacitative arms is determined using a coplanar capacitance calculator, the resonator frequency is measured and thus the impedance is determined to be $\sim$100~$\Omega$. The current fluctuations can then be analytically determined given the penetration depth of niobium being $\sim$50~nm~\cite{maxfield1965superconducting}.

\begin{figure*}
    \centering
    \includegraphics[width=\textwidth]{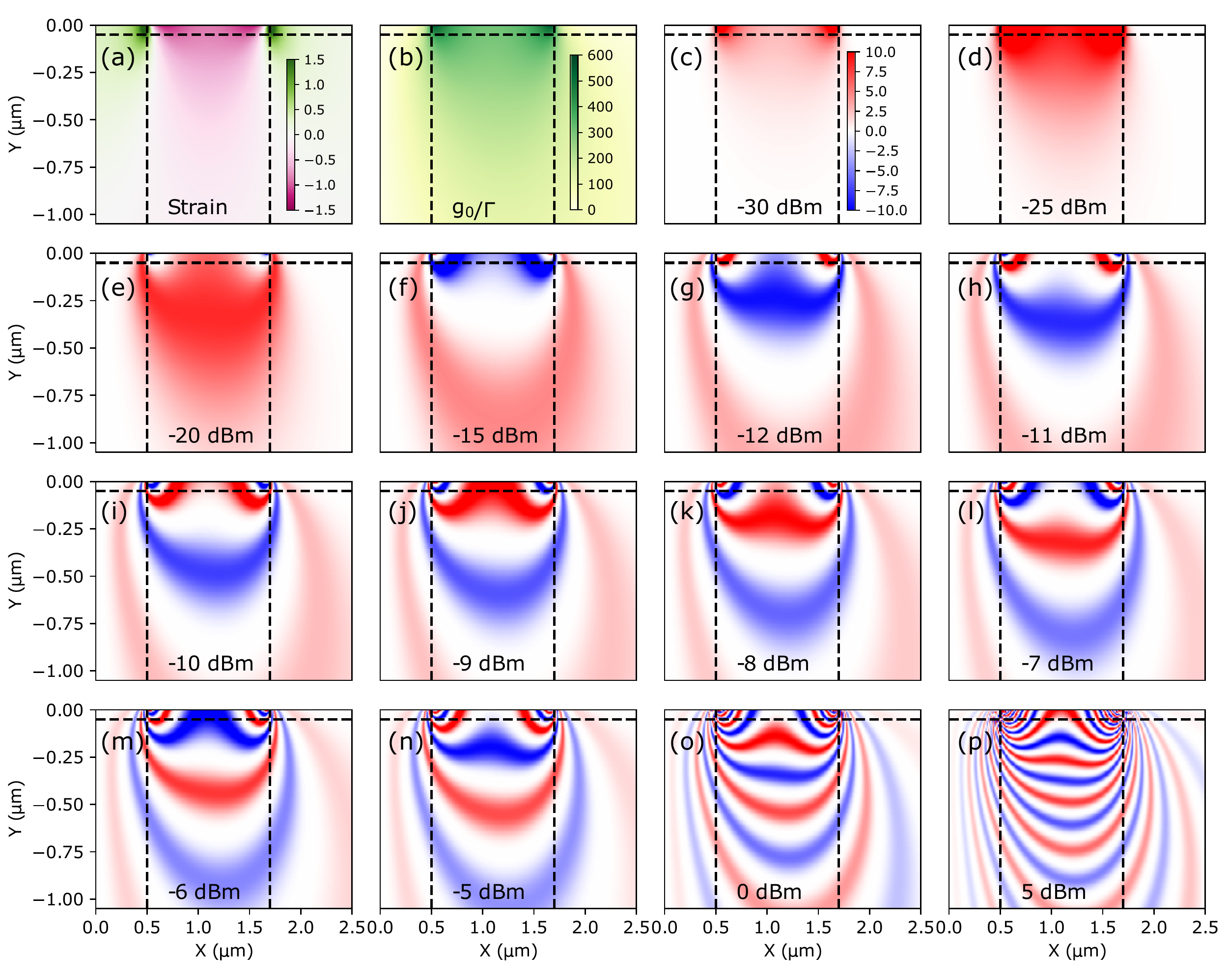}
    \caption{(a) Map of shifts in hyperfine constant (MHz) due to strain in the sample based on FE simulations as in Fig.~1(d). (b) A map of $g_0/\Gamma$ (Hz) for $S_{\rm z}$ transitions, a linear rescaling of $B_1$. (c-p) maps of contribution factor for $S_{\rm z}$ transitions $g_0\sin^3(\theta(\mathbf{r}))/\Gamma$ for different pulse powers where $\theta(\mathbf{r})$ has been calculated assuming $\Gamma = 1$. In all figures the horizontal dashed line is 50~nm below the surface, approximately where electrostatic calculations imply the depletion zone, where donors are ionised due to the Schottky barrier at the Nb/Si interface~\cite{pla2018strain}, ends. The vertical lines are at either side of the microresonator.}
    \label{fig:contributions_sz}
\end{figure*}

As described in the main text we calculate a spatially varying contribution factor from the driven microwave field $B_{\rm 1, driven} = B_{1}A_{\rm pulse}$ at each pixel in the FE simulation. This is shown for $S_{\rm z}$ transitions in Fig.~\ref{fig:contributions_sz} for different powers of initial pulse. At high powers banding occurs where spins in consecutive bands provide opposite contributions. 

\begin{figure*}
    \centering
    \includegraphics[width=\textwidth]{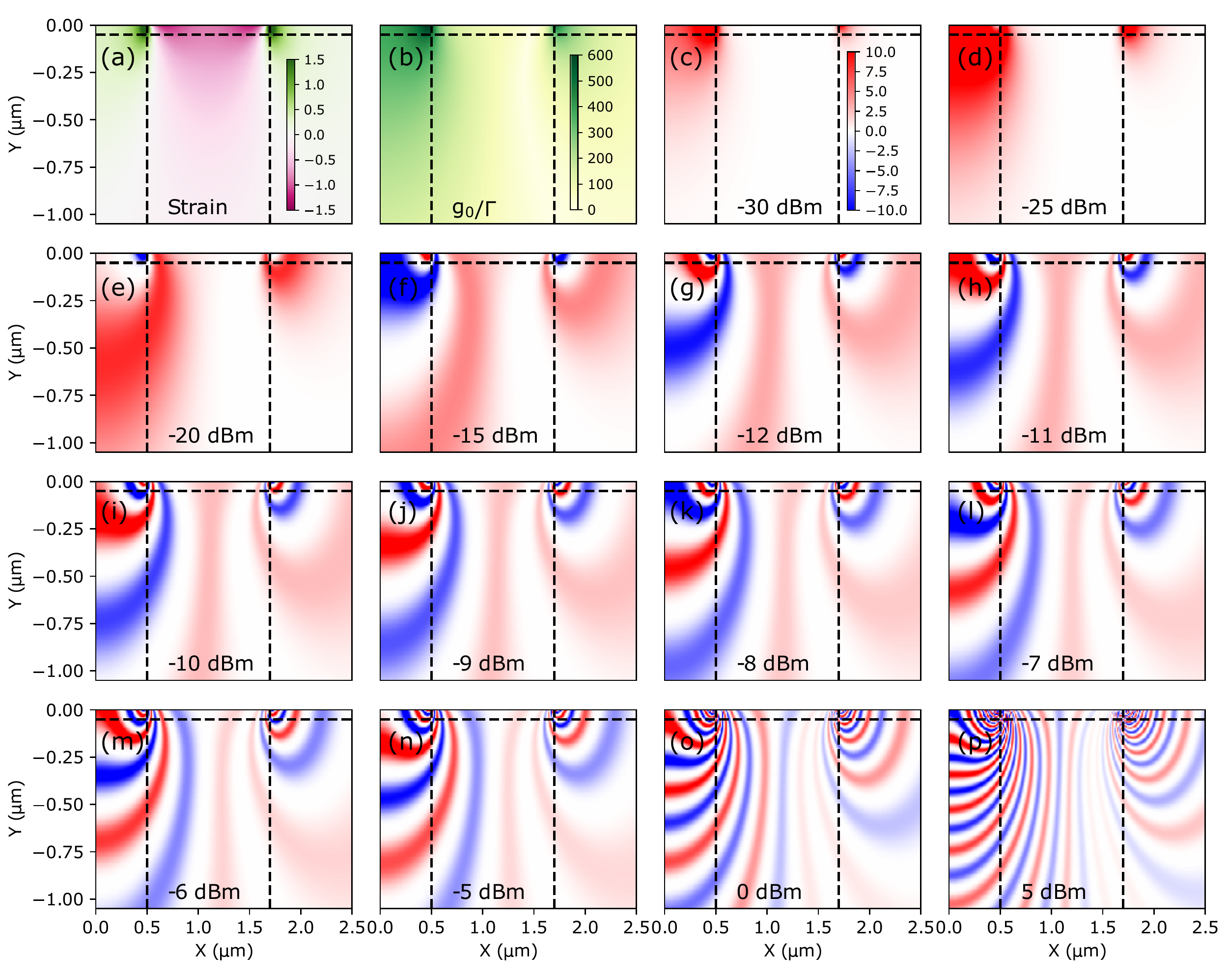}
    \caption{(a) Reproduction of \ref{fig:contributions_sz} for $S_{\rm x}$ transitions.}
    \label{fig:contributions_sx}
\end{figure*}

\begin{figure}[!htb]
    \centering
    \includegraphics[width=0.48\textwidth]{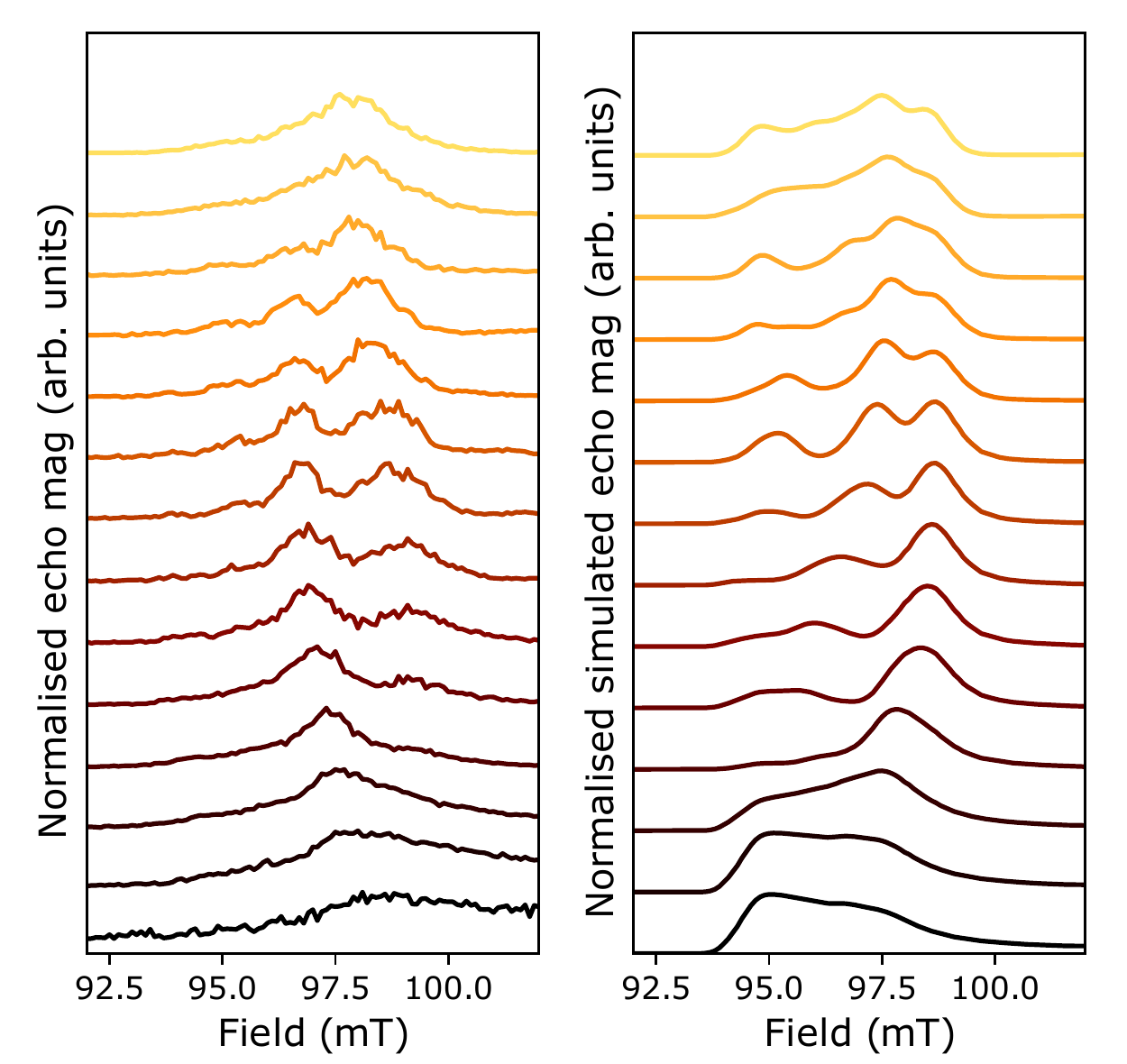}
    \caption{(a) CPMG EDFS reproduced from main text Fig.~4(g). (b) Calculated EDFS based on finite element simulations of strain and $B_1$ fields about the microresonator.}
    \label{fig:field_sweep_sim}
\end{figure}

Considering how strain shifts the hyperfine frequency, we can simulate an expected EDFS as is shown in Fig.~\ref{fig:field_sweep_sim}. This is achieved by solving the spin Hamiltonian for a number of different hyperfine constants. We then sum a series of Gaussians, one for each pixel in the FE simulation. The centre of the Gaussian is the frequency which the resonator crosses the transition based on the hyperfine constant at that pixel accounting for strain and mass shifts. The amplitude of the Gaussian is set by the contribution factor at that pixel. All Gaussians are given the same common width of 660~kHz. We do this for a few values of $A_{\rm pulse}$ and average them to account for inhomogeneities along the inductor wire. We then use the resonator dispersion to convert this frequency spectrum into field, thus simulating the EDFS. 

The simulated EDFSs have some broad qualitative agreement to the measured EDFSs. At low powers the line saturates in a broad peak. At high powers there is a peak of approximately the correct width in field. At intermediate powers there are oscillations in intensity within the field sweep. These simulated EDFS do not however accurately reproduce the experimental EDFS. There are many potential candidates for this discrepancy; significant, unaccounted, inhomogeneities along the inductor wire, contribution factors in the CPMG pulse sequence not following the $\sin^3(\theta)$ dependence of Hahn echo, errors in the FE simulation of strain, incorrect values of penetration depth being used to calculate $B_{\rm 1}$ to name a few. The broad qualitative agreement is promising, indicating that we capture the main features of this system. Further work, particularly on the strain simulations, is needed for a good qualitative agreement between simulation and experiment. 

We also consider, and rule out, the Meissner effect as the dominant cause of this line broadening. Using COMSOL, we model the resonator as strongly diamagnetic and compute the deformed static field about the resonator. Using this static field to compute the spin Hamiltonian results in EDFS drastically different to experiment and, most pertinently, means that at low powers the simulated peak position, which is already too low in field, shifts further downwards in field. These low power spectra, arising from spins closest to the resonator, are where this effect should be strongest and given that the incorrect sign of the effect, we conclude is not the dominant broadening process.

\subsection{Cooperativity and strong coupling in the small $\kappa$ limit}

The expression for cooperativity, $C=g_{\rm ens}^2/\kappa\gamma = g_0^2N/\kappa\gamma$, is derived from the Tavis-Cummings Hamiltonian considering a cavity coupled to $N$ two-level-systems --- in our case, spins. When the spins are inhomogeneously broadened such that the homogeneous linewidth, $\gamma_\text{h} = 1/T_2$, is small and we are in the regime $\gamma_\text{h} \ll \kappa<\gamma$, as is the case in this paper, these spins can fall outside of the cavity linewidth. The traditional definition of cooperativity indicates that spins outside the bandwidth of the cavity (i.e. no spectral overlap) contribute to the cooperativity.

In the limit $\gamma_\text{h} \ll \kappa<\gamma$, it makes more sense to consider only the spins within the resonator bandwidth, such that any change to spins that fall outside the resonator bandwidth cannot affect $C$ or the coupling. We can instead use the coupled spin linewidth $\gamma_\text{corr} \equiv \kappa$ and make a correction to the number of spins $N$ in the ensemble to ensure we only count spins within the resonator bandwidth. The corrected spin number is

\begin{equation}
N_\text{corr} = N \frac{\int_{-\infty}^{\infty} \frac{\eta}{\eta_\text{max}} \rho ~df}{ \int_{-\infty}^{\infty} \rho ~df}
\end{equation}
where $\rho$ is the spin spectral density, $\eta$ is the resonator power spectral density and $\eta_\text{max}$ is the maximum value of $\eta$, such that $\eta/\eta_\text{max}$ is the dimensionless resonator lineshape, normalised to an amplitude of 1. The term $\rho\eta/\eta_\text{max}$ is effectively a corrected spin lineshape that accounts only for spins coupled to the resonator. The ratio of the area of the corrected lineshape to the original lineshape gives the correction factor to $N$. Using this, we can now write the corrected ensemble cooperativity:
\begin{equation}
    C_\text{corr} = \frac{g_0^2 N_\text{corr}}{\kappa^2}
\end{equation}

Now consider the simple case of a uniformly broadened spin ensemble coupled to a narrow-bandwidth resonator with identical lineshape and with no detuning between the two. In this case the spin spectral density is approximately constant over the bandwidth of the resonator and the corrected spin lineshape takes on exactly the lineshape of the resonator, with the same amplitude as the original spin line. Then $N_\text{corr}$ is simply $N$ multiplied by the ratio of the resonator linewidth to the spin ensemble linewidth
\begin{equation}
N_\text{corr} = N \frac{\kappa}{\gamma}
\label{eq:Ncorr}
\end{equation}
which yields the original term for cooperativity:
\begin{equation}
    C_\text{corr} = \frac{g_0^2 N}{\kappa^2}\frac{\kappa}{\gamma} = \frac{g_0^2 N}{\kappa\gamma}
\end{equation}
where cooperativity continues to increase with narrowing resonator linewidth, despite the fact that this reduces the number of spins within the cavity.

In general the spin ensemble lineshape is \textit{not} identical to that of the resonator, nor is it uniformly broadened. Indeed, in these experiments the spin lineshape is asymmetric and modelled by three separate gaussian functions (six in the case of an $S_\text{x}$ transition), while the resonator is approximately Lorentzian. However this gives a good approximation and allows us to estimate the cooperativity in the narrow resonator limit in a simple manner.

Based on this, we can use the FE simulations of the resonator to estimate the ensemble coupling strength for a given distribution of spins. At each pixel the ratio of the single spin coupling, $g_0$, to the transition matrix element, $\Gamma$, is computed by 
$g_0/\Gamma = B_1\gamma_e$. When $B_1$ is inhomogeneous the ensemble coupling is given by $g_{\rm ens} = \sqrt{\sum{g_0^2}}$ where the sum runs over spins. Voxels are defined by the 2D pixels from the simulation with out-of-page extent given by the length of the inductor (assuming that current doesn't vary significantly along the length of the inductor). Each voxel has $N_{\rm Bi} = \rho_{\rm Bi}(\mathbf{r})V$ bismuth atoms within it where $\rho_{\rm Bi}(\mathbf{r})$ is the local density of bismuth atoms and $V$ is the voxel volume. $N_{\rm Bi}$ is adjusted by the thermal population and then, as above, corrected so that only spins inside the cavity linewidth couple to the resonator as in Eq.~\ref{eq:Ncorr}. For the transition shown in Fig.~2 main text this computation predicts an ensemble coupling strength of $\sim$260~kHz, which is a factor of $\sim 2$ larger than the experimental result. The discrepancy between experiment and simulation is likely due to a combination of uncertainties in the film kinetic inductance, the depletion zone depth, donor activation, a drop in current in the inductor close to the capacitive arms and spins close to the resonator having a greater linewidth and more often falling outside the cavity bandwidth. The variation in current along the inductor could be accounted for by additional simulations of the resonance mode using software packages such as CST Studio Suite.

Using these FE models we can show that continuing to add spins further from the resonator increases the coupling strength. We find that for resonators without the double-back inductor (i.e. a straight wire connecting capacitive arms) this effect is stronger as the $B_1$ fields are less confined to be local to the resonator. However, for a high fidelity memory, the spins must both add to cooperativity, and also be refocused in a quantum memory protocol. The weak $B_1$ fields means that refocusing will be inefficient and so this approach to increasing cooperativity is unsuitable if the intended application is for memories, or any application where coherent control of spins is required. 

The conditions for strong coupling should also be reconsidered. As we are interested in transferring information between the resonator and spin ensemble, being able to resolve a vacuum Rabi splitting is not essential. We are interested primarily in the loss rates from the resonator and spin ensemble, and as such the regime required for quantum memories is actually $g_0\sqrt N\gg \kappa,\gamma_\text{h}$. One can also define the \textit{homogeneous cooperativity}:
\begin{equation}
C_\text{h} = \frac{g_0^2 N}{\kappa\gamma_\text{h}}.
\end{equation}
The efficiency of the quantum memory asymptotically approaches 1 as $C_\text{h} \rightarrow \infty$ \cite{afzelius2013proposal}. This distinction between the homogeneous and inhomogeneous spin linewidths is why it is possible to operate an efficient quantum memory without being in the traditionally defined strong coupling regime, $g_0\sqrt N\gg \kappa, \gamma$. The coupling of a narrow cavity to an inhomogeneously broadened spin ensemble has also been discussed in Refs~[\citenum{grezes2016towards, grezes2015towards, julsgaard2012reflectivity, afzelius2013proposal}].

\subsection{Calculation of resonator photon number}
The Rabi oscillations obtained in Fig.4 enable us to estimate the number of photons in the cavity. Using the Tavis-Cummings model for an ensemble of spins coupled to a cavity, the Rabi frequency $\Omega$ of the spins when the resonator contains $n$ photons is given by:

\begin{equation}
\Omega = 2 g_0 \sqrt{n}
\label{eq:Rabi}
\end{equation}

We can calculate $g_0$, assuming that $T_1$ relaxation is dominated by the Purcell effect\cite{bienfait2016controlling}, using the equation:

\begin{equation}
\frac{1}{T_1} = \frac{4 g_0^2}{\kappa}
\label{eq:Purcell}
\end{equation}

Where $\kappa$ is the resonator half-width and $g_0$ is the resonator to single spin coupling rate. As we have seen, in this system $g_0$ varies greatly according to the position of the spins relative to the resonator and as such it is not possible to get a value for $T_1$ of the entire ensemble. We take a rough estimate of $T_1 \approx 1.5$~s to calculate $g_0\approx350~$Hz for spins close to the 6764~MHz resonator. We then use Eq.\ref{eq:Rabi} assuming $\Omega = 500$~kHz (the duration of the pre-rotation pulse used in Fig.4 was 2~$\upmu$s) to calculate a photon number of $\sim5\times 10^5$. Note that this is only a very rough estimate and applies only to low power measurements of the spins closest to the resonator.

\bibliography{bibliography}